\documentclass[twocolumn,showpacs,showkeys,preprintnumbers,amssymb,aps,superscriptaddress,pre]{revtex4-2}
\usepackage{graphicx}% Include figure files
\usepackage{dcolumn}% Align table columns on decimal point
\usepackage{bm}% bold math
\usepackage{epsfig}
\usepackage{color}
\usepackage{amsmath}
\usepackage{hyperref}
\hypersetup{colorlinks=true,linkcolor=blue,urlcolor=blue,citecolor=blue}

\begin{document}

\title{Magnetoelastic signatures of thermal and quantum phase transitions in a deformable Ising chain under a longitudinal and transverse magnetic field}
\author{D\'avid Siv\'y} 
\affiliation{Institute of Physics, Faculty of Science, P. J. \v{S}af\'{a}rik University, Park Angelinum 9, 04001 Ko\v{s}ice, Slovakia}
\author{Jozef Stre\v{c}ka}
\email{jozef.strecka@upjs.sk}
\affiliation{Institute of Physics, Faculty of Science, P. J. \v{S}af\'{a}rik University, Park Angelinum 9, 04001 Ko\v{s}ice, Slovakia}

\date{\today}

\begin{abstract}
We investigate a deformable spin-1/2 Ising chain subjected to either a longitudinal or a transverse magnetic field, which incorporates a magnetoelastic coupling linearly dependent on a lattice distortion parameter. Within the harmonic and static adiabatic approximations, the variational Gibbs free energy is evaluated exactly using transfer-matrix and Jordan-Wigner fermionization techniques and then minimized self-consistently with respect to the lattice distortion parameter. This approach enables a unified description of magnetic and elastic properties including the magnetization, magnetic susceptibility, lattice distortion, inverse compressibility, and relative change in the sound velocity. In a longitudinal magnetic field, the deformable Ising chain displays a line of discontinuous thermal phase transitions terminating at a critical point. The discontinuous transitions are accompanied by metastable states, which give rise to a hysteresis loop at low temperatures. In contrast, the deformable Ising chain in a transverse field undergoes exclusively a continuous quantum phase transition at zero temperature with no indication of thermal phase transitions. The magnetic susceptibility and inverse compressibility exhibit cusp- and dip-like anomalies at discontinuous phase transitions, while a diverging susceptibility and vanishing inverse compressibility characterize the continuous phase transitions. An elastic softening of the deformable chain near thermal and quantum phase transitions manifest itself also through a significant sound attenuation. 
\end{abstract}
 

\maketitle

\section{Introduction}

One-dimensional (1D) quantum spin systems attract considerable attention in statistical and condensed-matter physics, because their relatively simple structure often allows for exact analytical solutions despite the strong correlations inherent to quantum many-body problems \cite{bet31,lie61,mat93,fra17}. These exactly solvable models provide valuable insight into a wide range of quantum phenomena including quantum phase transitions, critical behavior, entanglement, and low-dimensional magnetism \cite{str14,roj17,zhe19,fre19,gal21,mic21,zhe22,nis23}. Owing to their analytical tractability, they serve as important benchmark for exploring fundamental aspects of quantum many-body physics and for testing various theoretical and numerical methods. Moreover, the understanding gained from exactly solvable quantum spin chains often provides useful starting point and/or guidance for the study of more complex 1D quantum spin systems, where exact solutions are generally not available \cite{ver16,kar19,ver19,sou20,ver21,str22}.

Although 1D spin systems with purely Ising exchange interactions may at first sight appear oversimplified, they play an important role due to the possibility of obtaining exact analytical solutions. In particular, the spin-1/2 Ising chain in a longitudinal magnetic field can be solved exactly by the transfer-matrix method \cite{isi25,kra41}, while the spin-1/2 Ising chain in a transverse field admits an exact solution through the Jordan–Wigner fermionization technique \cite{lie61,kat63,pfe70}. These exact results make the Ising model one of the fundamental paradigms for studying collective behavior and quantum phase transitions in low-dimensional spin systems. Moreover, the study of Ising spin systems is not merely of theoretical interest as several real magnetic materials exhibit strong Ising anisotropy \cite{wol00,jon01}. Prominent examples of quasi-one-dimensional Ising-chain compounds include BaCo$_2$V$_2$O$_8$ \cite{yam11}, CoCl$_2\cdot2$D$_2$O \cite{lar17}, SrCo$_2$V$_2$O$_8$ \cite{cui19}, RbCoCl$_3$ \cite{men20}, and CoNb$_2$O$_6$ \cite{hau24}. These systems provide valuable experimental realizations of Ising-like physics and stimulate further theoretical investigations of 1D spin models.

The famous nonexistence theorems state that 1D spin systems with short-range interactions and nonsingular potentials cannot exhibit thermal phase transitions at finite temperatures \cite{van50,cue04}. Nevertheless, recent studies of certain 1D Ising spin systems have demonstrated extremely sharp crossovers closely resembling true thermal phase transitions and these intriguing phenomena have been termed as pseudo-transitions \cite{str20,hut21,szn22,yin24a,yin24b,yas24,roj26}. Importantly, the nonexistence theorems do not exclude the possibility of quantum phase transitions, which occur strictly at zero temperature and are driven by quantum fluctuations upon varying external parameters such as magnetic field or pressure \cite{sac11a}. Quantum phase transitions in 1D spin systems continue to attract considerable attention \cite{zha24,ish24,xia24,sac24,rib24}, since they strongly influence magnetic and thermodynamic properties even at finite temperatures where their signatures still remain experimentally observable \cite{bal10,sac11b}. In particular, the proximity to a quantum critical point may lead to pronounced elastic anomalies including lattice softening and enhanced sound attenuation, which provide another experimentally accessible signatures of quantum phase transitions at finite temperatures \cite{yam11,wan18,wol04,mat20,hau24}.

To make 1D quantum spin models more realistic, it is thus quite natural to extend the framework of rigid lattices by incorporating magnetoelastic coupling, which allows the lattice spacing to adjust in response to the magnetic state. Within this approach the lattice deformation is typically treated in the harmonic approximation, while the magnetic subsystem is handled exactly. The interplay between spin and lattice degrees of freedom can give rise to a variety of intriguing phenomena. A prominent example is the magnetostructural spin-Peierls transition, which involves spontaneous lattice dimerization driven by spin–phonon coupling \cite{pin71,pyt74}. This effect has been experimentally confirmed in several 1D magnetic compounds such as the inorganic spin-chain material CuGeO$_3$ \cite{has93} and the organic verdazyl-radical salt \cite{yam23}. Apart from dimerization, magnetoelastic coupling may also induce a uniform lattice deformation, which modifies the magnetic interactions and can significantly affect the magnetic and elastic behavior of the system. From a theoretical perspective, quantum spin chains with magnetoelastic coupling represent a more involved problem than their perfectly rigid counterparts though several models remain amenable to rigorous analysis when the magnetic subsystem is solved exactly and the lattice degrees of freedom are treated within the harmonic approximation. Notable examples include the spin-1/2 XX chain in a transverse magnetic field \cite{ori05}, the spin-1/2 Ising chain in longitudinal or transverse magnetic fields \cite{der13}, as well as various Ising–Heisenberg bond-alternating chains \cite{der13}.

In the present work, we investigate magnetic and elastic properties of a deformable spin-$1/2$ Ising chain subjected to either a longitudinal or a transverse magnetic field at finite temperatures. In both cases, magnetoelastic coupling is incorporated through a uniform lattice distortion treated within the harmonic approximation in a static regime. The spin-$1/2$ Ising chain in a longitudinal magnetic field is solved exactly using the transfer-matrix method \cite{isi25,kra41}, whereas the spin-$1/2$ Ising chain in a transverse field is treated by means of the Jordan–Wigner fermionization technique \cite{lie61,kat63,pfe70}. The exact treatment of the magnetic subsystem provides the variational Gibbs free energy, which is subsequently minimized with respect to the lattice distortion parameter in a self-consistent manner. Within this framework, we analyze the dependence of the magnetization, magnetic susceptibility, lattice distortion parameter, inverse compressibility, and relative change in the sound velocity on the magnetic field, pressure, and temperature. While deformable Ising chains were previously examined at zero temperature \cite{der13}, the present study extends this analysis to finite temperatures and provides a comprehensive description of both thermal and quantum critical behavior arising from the magnetoelastic coupling.

The paper is organized as follows. In Sec.~\ref{model}, we introduce the deformable spin-$1/2$ Ising chain subjected to either a longitudinal or a transverse magnetic field and briefly outline the methods used to obtain the exact solutions. Section~\ref{results} presents and discusses the main results for the magnetization, magnetic susceptibility, lattice distortion parameter, inverse compressibility, and relative change in the sound velocity. Finally, Sec.~\ref{conclusion} summarizes the key findings.

\section{Model and methods}
\label{model}

To begin, we consider a deformable spin-1/2 Ising chain consisting of $N$ spins arranged 1D lattice and described by the Hamiltonian:
\begin{eqnarray}
{\cal \hat{H}} = \sum_{i=1}^{N}\left[J \left(\cos{\gamma} \hat{S}^{z}_{i}\hat{S}^{z}_{i+1} + \sin{\gamma} \hat{S}^{x}_{i}\hat{S}^{x}_{i+1} \right) - h\hat{S}_{i}^{z} \right]\!,
\label{generalHam}
\end{eqnarray} 
where $\hat{S}^{x}_{i} \equiv \frac{1}{2} \hat{\sigma}^{x}_{i}$ and $\hat{S}^{z}_{i} \equiv \frac{1}{2} \hat{\sigma}^{z}_{i}$ are two spatial components of the spin-1/2 operator assigned to the $i$-th lattice site ($i=1,\ldots,N$) expressed in terms of the Pauli matrices ($\hbar$ is set to unity). The parameter $J$ denotes the nearest-neighbor exchange coupling, and $h$ determines the strength of an external magnetic field applied along the $z$ axis. The parameter $\gamma$ controls a relative orientation of the magnetic field with respect to a quantization axis, the Ising spins are put into an external longitudinal (transverse) magnetic field for $\gamma=0$ ($\gamma=\pi/2$). In contrast to the standard rigid-lattice formulation, we assume that the 1D lattice is deformable and the overall lattice length depends on the applied (dimensionless) pressure $p$. The lattice constant $a_{0}$ is assigned to an equilibrium lattice spacing between the nearest-neighbor spins, whereby a uniform change of the lattice constant $a = a_0 (1 + \delta)$ can be characterized by a dimensionless lattice distortion parameter $\delta$. A compressive (tensile) strain $\delta<0$ ($\delta>0$) corresponds to a compressive (tensile) stress $p>0$ ($p<0$). Within the harmonic approximation, the change in lattice spacing associated with a uniform deformation results in the total elastic energy penalty $N\alpha\delta^2/2$, where $\alpha$ denotes the elastic lattice constant. In the following, we adopt the static approximation in the adiabatic regime \cite{ori05,der13}, which accounts in the Hamiltonian (\ref{generalHam}) for the uniform change of exchange-coupling constant $J \equiv J(a)$ from its equilibrium value $J_{0} \equiv J(a_0)$: 
\begin{eqnarray}
J=J_{0}\left(1-|\kappa| \delta \right).
\label{Jdef}
\end{eqnarray}
This change results from the lattice strain, which depends on the magnetoelastic coupling constant $\kappa = \frac{1}{J_0} (\frac{\partial J}{\partial a})_{a=a_0}<0$. Note furthermore that only very small lattice distortions $\delta \ll 1$ are available within the harmonic and adiabatic approximations. 

To proceed further we evaluate the variational Gibbs free energy per spin involving three distinct contributions: 
\begin{eqnarray}
g(T,h,p;\delta) = \frac{1}{2}\alpha\delta^2 + p\delta + g_m(T,h;\delta).
\label{gibbs1}
\end{eqnarray}
The first term represents the elastic energy in the harmonic approximation, the second term accounts for the work done by the external pressure, and the magnetic contribution $g_{m}$ can be calculated in the thermodynamic limit $N \to \infty$ according to the equation:
\begin{eqnarray}
g_{m}(T,h;\delta) = -\lim_{N\to\infty} \frac{1}{N}k_{\rm B}T {\rm ln} Z.
\label{MagGibbs1}
\end{eqnarray}      
Here, $k_{\rm B}$ denotes the Boltzmann constant, $T$ is the absolute temperature, and $Z = {\rm Tr}~ {\rm exp}(-\beta \mathcal{\hat{H}})$ is the partition function with $\beta=1/(k_{\rm B}T)$. Having calculated the magnetic contribution to the Gibbs free energy (\ref{MagGibbs1}), one can derive an equation of state by minimizing the overall variational Gibbs free energy (\ref{gibbs1}) with respect to the distortion parameter $\delta$. This minimization procedure provides an equilibrium value of the distortion parameter $\delta(T,h,p)$ satisfying the condition \cite{ori05,der13}:
\begin{eqnarray}
\delta(T,h,p)= -\frac{1}{\alpha}\left[p+\left( \frac{\partial g_{m}}{\partial \delta} \right)_{\delta = \delta(T,h,p)}\right].
\label{delta1}
\end{eqnarray}
From the equilibrium value of the lattice distortion parameter $\delta(T,h,p)$ given by Eq. (\ref{delta1}) one may subsequently easily calculate the (inverse) lattice compressibility:
\begin{eqnarray}
\frac{1}{ \varkappa} \equiv - \frac{\partial \delta(p)}{ \partial p},
\label{invC1}
\end{eqnarray}
which characterizes the lattice response to the applied pressure. In addition, the lattice compressibility is directly related to the experimentally accessible sound velocity, which is in the long-wavelength limit proportional to:
\begin{eqnarray}
c(h) \propto a \sqrt{\varkappa(h) / m_0},
\label{soundh}
\end{eqnarray}
where $m_0$ denotes the mass of the spin-1/2 magnetic ion. Consequently, the relative change of the sound velocity induced by the external magnetic field can be estimated from the formula:
\begin{align}
\frac{\Delta c}{c_0} \! &\equiv \! \frac{c(h) \!-\! c(0)}{c(0)} \!\approx\! \frac{\sqrt{\varkappa(h)} \!-\! \sqrt{\varkappa(0)}}{\sqrt{\varkappa(0)}} \!=\! \sqrt{ \frac{\varkappa(h)}{\alpha} } \!-\! 1,
\label{deltaSound}
\end{align}
which is based on the assumption that variations in the lattice spacing $a$ are negligible compared to that ones in the exchange coupling $J$. In the absence of magnetic contribution to the variational Gibbs free energy  one recovers the value $\varkappa(0) = \alpha$ corresponding to a purely harmonic nonmagnetic lattice \cite{der13}.

Finally, one needs to evaluate the full variational Gibbs free energy in order to derive magnetic quantities such as the magnetization and magnetic susceptibility. This is accomplished by substituting the equilibrium value of the lattice distortion parameter given by Eq. (\ref{delta1}) into the variational Gibbs free energy (\ref{gibbs1}). In the present work, we particularly focus on the system at finite temperatures, since the zero-temperature case has already been analyzed in our previous study (see Ref. \cite{der13}). To this end, the key remaining step thus represents the evaluation of the magnetic contribution to the variational Gibbs free energy, which determines the magnetic and elastic behavior of the system at finite temperatures.

\subsection*{A. Longitudinal field}
\label{subsection_A}

Let us at first determine the magnetic contribution to the variational Gibbs free energy for the deformable spin-1/2 Ising chain in a longitudinal magnetic field given by the total Hamiltonian (\ref{generalHam}) for $\gamma=0$, which can be rewritten into the most symmetric form as follows:
\begin{eqnarray}
{\cal H} = \sum_{i=1}^{N}\left[J_{0}\left(1-|\kappa|\delta\right)S^{z}_{i}S^{z}_{i+1} - \frac{h}{2}\left(S_{i}^{z} + S_{i+1}^{z}\right) \right].
\label{generalHamLong}
\end{eqnarray}
Inserting the Hamiltonian (\ref{generalHamLong}) into the partition function yields its factorized form:
\begin{eqnarray}
Z =  \sum_{\left\{S_{i}^{z}\right\}}{\rm exp}\left(-\beta \mathcal{H}\right) = \sum_{\left\{S_{i}^{z}\right\}} \prod_{i=1}^{N} {\rm \textbf{T}}(S_{i}^{z},S_{i+1}^{z}). 
\label{partIsi1}
\end{eqnarray}
The expression ${\rm \textbf{T}}(S_{i}^{z},S_{i+1}^{z})$ can be identified with a two-by-two transfer matrix:
\begin{eqnarray}
{\rm \textbf{T}}(S_{i}^{z},S_{i+1}^{z}) \!=\!
\begin{pmatrix}
{\rm T}_{++}\left(+\frac{1}{2},+\frac{1}{2}\right) & {\rm T}_{+-}\left(+\frac{1}{2},-\frac{1}{2}\right) \\
{\rm T}_{-+}\left(-\frac{1}{2},+\frac{1}{2}\right) & {\rm T}_{--}\left(-\frac{1}{2},-\frac{1}{2}\right)
\end{pmatrix}\!.
\label{transIsi2}
\end{eqnarray}
whose individual matrix elements are defined by:
\begin{eqnarray}
{\rm \textbf{T}}(S_{i}^{z},S_{i+1}^{z}) \!=\! {\rm exp} \! \left[-\beta J S_{i}^{z} S_{i+1}^{z} \!+\! \frac{\beta h}{2}\left(S_{i}^{z} \!+\! S_{i+1}^{z}\right)\!\right]\!\!. 
\label{transIsi1}
\end{eqnarray}
The summation over spin states is equivalent to a multiplication of the transfer matrices and this allows to express the partition function (\ref{partIsi1}) 
within the transfer-matrix method in the following form:
\begin{eqnarray}
Z \!=\! \sum_{S_{1}^{z}=\pm\frac{1}{2}}\!{\rm \textbf{T}}^{N}(S_{1}^{z},S_{1}^{z})
={\rm Tr}~{\rm \textbf{T}}^{N} = \rm{\lambda}_{+}^{N} + \rm{\lambda}_{-}^{N},
\label{PartFinal1}
\end{eqnarray} 
where $\lambda_{\pm}$ represent two eigenvalues of the transfer matrix (\ref{transIsi2}) obtained after its diagonalization:
\begin{eqnarray}
\lambda_{\pm} \!=\! \frac{1}{2}\! \left[{\rm T}_{++} \!+\! {\rm T}_{--} \pm \sqrt{({\rm T}_{++} \!-\! {\rm T}_{--})^{2} \!+\! 4 {\rm T}_{+-}^{2}} \right]\!.
\label{TransEigenVal1}
\end{eqnarray}
In the thermodynamic limit $N \to \infty$, the magnetic contribution to the variational Gibbs free energy (\ref{MagGibbs1})
in turn follows from the largest eigenvalue of the transfer matrix:
\begin{eqnarray}
g_{m}=-\lim_{N\to\infty} \frac{1}{N}k_{\rm B}T {\rm ln} Z = -k_{\rm B} T \ln \lambda_{+}.
\label{MagGibbs2}
\end{eqnarray} 
To obtain the equation of state, the magnetic contribution to the Gibbs free energy (\ref{MagGibbs2}) is inserted into the transcendental equation for the distortion parameter (\ref{delta1}):
\begin{eqnarray}
\delta(T,h,p) &=& -\frac{1}{\alpha}\left[ p - \frac{1}{4}J_{0}|\kappa| + \frac{J_{0}|\kappa|}{2\lambda_{+}} \right. \nonumber \\
&\times& \left. \frac{\exp\left[\frac{3}{4}\beta J_{0}\left(1\!-\! |\kappa|\delta\right)\right]}{\sqrt{\sinh^{2}\left(\frac{\beta h}{2}\right)+\exp\left[\beta J_{0}\left(1\!-\! |\kappa|\delta \right)\right]}} \right]\!\!,
\label{deltaFinalLong}
\end{eqnarray}
which should be solved in a self-consistent manner. Using the definition (\ref{invC1}) and the obtained equation of state (\ref{deltaFinalLong}) the inverse compressibility reads
\begin{eqnarray}
\varkappa = \alpha + \beta\left( \alpha\delta + p - \frac{1}{4}J_0 |\kappa| \right) \Bigg\{ \alpha\delta + p + \frac{3}{4}J_0 |\kappa|
\nonumber \\
- \frac{1}{2}\frac{J_0 |\kappa| \exp\left[\beta J_0 (1 - |\kappa|\delta)\right]}{\sinh^2\left(\frac{\beta h}{2}\right) + \exp\left[\beta J_0 (1 - |\kappa|\delta)\right]} \Bigg\}.
\label{invC2}
\end{eqnarray}
Last but not least, the magnetization per spin can be obtained using the standard relation from the variational Gibbs free energy (\ref{MagGibbs2}) by using the fact that 
$\partial g / \partial \delta = 0$:
\begin{eqnarray}
m = \frac{1}{2}\frac{\sinh\left(\frac{\beta h}{2}\right)}{\sqrt{\sinh^2\left(\frac{\beta h}{2}\right) + \exp\left[\beta J_0 (1 - |\kappa|\delta)\right]}}.
\label{LongMagExact}
\end{eqnarray}
Similarly, the magnetic susceptibility per spin reads:
\begin{eqnarray}
\chi = \frac{\partial m(h,\delta(h))}{\partial h} = \frac{\partial m(h)}{\partial h} + \frac{\partial m(\delta)}{\partial \delta}\frac{\partial \delta(h)}{\partial h}.
\label{LongSusExact}
\end{eqnarray}

\subsection*{B. Transverse field}
\label{subsection_B}

The total Hamiltonian (\ref{generalHam}) for the particular choice of $\gamma=\pi/2$ reduces to the deformable spin-1/2 Ising chain in a transverse magnetic field: 
\begin{eqnarray}
{\cal \hat{H}} = \sum_{i=1}^{N}\left[J_{0}\left(1-|\kappa|\delta\right)\hat{S}^{x}_{i}\hat{S}^{x}_{i+1} - h\hat{S}_{i}^{z} \right].   
\label{IsingGenT}
\end{eqnarray}
To diagonalize the Hamiltonian (\ref{IsingGenT}) and determine the magnetic contribution to the variational Gibbs free energy (\ref{MagGibbs1}), one can adopt a series of transformations. First, we start by rewriting two spatial components of the spin operator $\hat{S}_{j}^{\alpha}$ ($\alpha=x,z$) assigned to $j$-th lattice site using spin ladder operators $\hat{S}_{j}^{\mp} = \hat{S}_{j}^{x} \mp {\rm i} \hat{S}_{j}^{y}$ in terms of which the Hamiltonian (\ref{IsingGenT}) acquires the form:
\begin{eqnarray}
{\cal \hat{H}} &=& \frac{J_{0}}{4} (1-|\kappa|\delta) \sum_{i=1}^{N-1}\left( 
\hat{S}_{i}^{+}\hat{S}_{i+1}^{+} + \hat{S}_{i}^{-}\hat{S}_{i+1}^{-} \right. \nonumber \\
&& \left. + \hat{S}_{i}^{+}\hat{S}_{i+1}^{-} + \hat{S}_{i}^{-}\hat{S}_{i+1}^{+} 
\right) \!-\! h \sum_{i=1}^{N} \left( \hat{S}_{i}^{+}\hat{S}_{i}^{-} \!-\! \frac{1}{2} \right)\!.
\label{transHamOpen}
\end{eqnarray}
For simplicity, we have changed the boundary conditions from periodic to open ones by noting that in the thermodynamic limit this change will not affect the final result for the variational Gibbs free energy. The Hamitonian (\ref{transHamOpen}) can be further re-expressed in terms of Fermi operators introduced via the Jordan-Wigner transformation:
\begin{eqnarray}
\hat{c}_{1} &= \hat{S}_{1}^{-}\!; \,\, \hat{c}_{i} &= \! \left[ \prod_{j=1}^{i-1}\left( -2\hat{S}_{j}^z \right) \right] \! \hat{S}_{i}^{-}\!, \, \! (i = 2,3,\ldots,N).
\label{newFerm}
\end{eqnarray}
The newly defined operators (\ref{newFerm}) and their Hermitian conjugates ($\hat{c}^{\dagger}_{1},\hat{c}^{\dagger}_{i}$) indeed satisfy fermionic anticommutation relations and the Hamiltonian (\ref{transHamOpen}) can be consequently rewritten in terms of Fermi creation and annihilation operators as follows:
\begin{eqnarray}
\hat{{\cal H}} &=& \frac{J_{0}}{4} (1-|\kappa|\delta)\sum_{i=1}^{N-1}\left(\hat{c}_{i}^{\dagger} \hat{c}_{i+1}+ \hat{c}_{i}^{\dagger} \hat{c}_{i+1}^{\dagger} + \text{h.c.}  \right) \nonumber \\
&-& h \sum_{i=1}^{N} \left( \hat{c}_{i}^{\dagger} \hat{c}_{i} - \frac{1}{2}  \right).
\label{transHamJord}
\end{eqnarray}
To diagonalize the Hamiltonian (\ref{transHamJord}) we continue with the discrete Fourier transformation:
\begin{eqnarray}
\hat{c}_{j} = \frac{1}{\sqrt{N}}\sum_{k}{\rm e}^{{\rm i}kj}\hat{d}_{k},
\label{fourierTran}   
\end{eqnarray}
where $k = \lambda\frac{2\pi}{N}$ denotes the momentum of a spinless fermion with $\lambda = -\frac{N}{2},-\frac{N}{2}+1,\cdots,\frac{N}{2}-1$. Applying the discrete Fourier transformation (\ref{fourierTran}), the Hamiltonian (\ref{transHamJord}) is partially diagonalized:
\begin{eqnarray}
\mathcal{\hat{H}} &=& \sum_{k=0}^{\pi} \bigg\{ \! \left[\frac{J_{0}}{2} (1-|\kappa|\delta) \cos k - h\right] \! 
\left(\hat{d}_{k}^{\dagger} \hat{d}_{k} + \hat{d}_{-k}^{\dagger} \hat{d}_{-k} \right) \nonumber \\ 
&& + {\rm i} \frac{J_{0}}{2} (1-|\kappa|\delta) \sin k \! \left( \hat{d}_{k}^{\dagger}\hat{d}_{-k}^{\dagger} 
\!+\! \hat{d}_{k}\hat{d}_{-k} \! \right) \!\! \bigg\} \!+\!\frac{N h}{2},
\label{forHam3}
\end{eqnarray}
which can be further simplified into a matrix form:
\begin{eqnarray}
\mathcal{\hat{H}} &=& \sum_{k=0}^{\pi} 
\begin{pmatrix}
\hat{d}_{k}^{\dagger} & \hat{d}_{-k} 
\end{pmatrix}
\begin{pmatrix}
A_{k} & {\rm i}B_{k} \\
-{\rm i}B_{k} & -A_{k}
\end{pmatrix} 
\begin{pmatrix}
\hat{d}_{k} \\ 
\hat{d}_{-k}^{\dagger} 
\end{pmatrix}, 
\label{forHamfinal}
\end{eqnarray}
with $A_{k}\!=\!\frac{J_{0}}{2}(1\!-\!|\kappa|\delta)\cos k \!-\!h$ and $B_{k} \!=\!\frac{J_{0}}{2}(1\!-\!|\kappa|\delta)\sin k$. To complete the diagonalization we apply the Bogoliubov transformation $\hat{\eta}_{k} = x_{k}\hat{d}_{k}+y_{k}\hat{d}_{-k}^{\dagger}$ ($x_{k}$ and $y_{k}$ are complex tranformation constants fulfilling the condition $\left|x_{k}\right|^2+\left|y_{k}\right|^2=1$), which brings the Hamiltonian (\ref{forHamfinal}) of the deformable spin-1/2 Ising chain in the transverse magnetic field into a fully diagonal form:
\begin{eqnarray}
\mathcal{\hat{H}}= \sum_{k=-\pi}^{\pi} \!\! \Lambda_{k}\left( \hat{\eta}_{k}^{\dagger}\hat{\eta}_{k} - \frac{1}{2} \right) 
= \sum_{k=-\pi}^{\pi} \!\! \Lambda_{k}\left(\hat{n}_{k} - \frac{1}{2} \right)\!, 
\label{bogoFinal}
\end{eqnarray}
where $\Lambda_{k}=\sqrt{A_{k}^{2}+B_{k}^{2}}$ is an elementary excitation spectrum and $\hat{n}_{k}=\hat{ \eta}_{k}^{\dagger}\hat{\eta}_{k}$ is the fermionic number operator with two eigenvalues $n_{k}=0,1$. Inserting the diagonal form of the Hamiltonian (\ref{bogoFinal}) into the partition function one obtains:
\begin{eqnarray}
Z &=& {\rm Tr}~ {\rm exp}\left(-\beta \mathcal{\hat{H}}\right) = \sum_{\left\{n_{k}\right\}}{\rm exp}\left[-\beta \sum_{k= -\pi}^{\pi}\Lambda_{k}\left( n_{k} - \frac{1}{2} \right)\right] \nonumber \\
&=& \prod_{k=-\pi}^{\pi}2\cosh \left( \frac{\beta \Lambda_{k}}{2} \right).
\label{finalMarks}
\end{eqnarray}
The magnetic part of the variational Gibbs free energy (\ref{MagGibbs1}) can be obtained from the partition function (\ref{finalMarks}) as:
\begin{eqnarray}
g_{m} \!=\! -k_{\rm B}T \ln{2} \!-\! \frac{k_{\rm B}T}{2\pi}\!\int_{-\pi}^{\pi}\!\! dk\ln{\left[\cosh{\left(\frac{\beta \Lambda_{k}}{2}\right)}\right]}\!.
\label{finalMarks3}
\end{eqnarray}
The equation of state in the form of a transcendent equation for the distortion parameter can be then obtained by inserting the variational Gibbs free energy (\ref{finalMarks3}) into Eq. (\ref{delta1}):
\begin{eqnarray}
\!\! \delta = \! - \frac{p}{\alpha} \!-\! \frac{J_{0}|\kappa|}{8\alpha\pi} \!\!\! \int_{-\pi}^{\pi} \!\!\!\!\!\! 
dk \frac{A_{k}\cos k \!+\! B_{k}\sin k}{\Lambda_{k}} \! \tanh \!\left(\!\frac{\beta\Lambda_{k}}{2}\!\!\right)\!\!.
\label{finalTransdelta}
\end{eqnarray}
Note that the equation of state (\ref{finalTransdelta}) should be solved in a self-consistent manner as the distortion parameter $\delta$ enters into the parameters $A_{k}$, $B_{k}$, and $\Lambda_{k}$. Using the equation (\ref{finalTransdelta}) and the definition (\ref{invC1}) allows one to calculate the inverse compressibility:
\begin{eqnarray}
\varkappa &=& \alpha + \frac{(J_0 |\kappa|)^2}{8\pi} \! \int_0^\pi \!\!\! dk \Bigg\{ \frac{1}{\Lambda_k^3}(A_k \! \cos k + B_k \! \sin k)^2 \\
&\times&\!\! \left[\! \tanh\!\left(\!\frac{\beta \Lambda_k}{2}\!\right) \!-\! \frac{\beta}{2}\Lambda_k \text{sech}^2\!\left(\!\frac{\beta \Lambda_k}{2}\!\right) \!\right] \!-\! \frac{1}{\Lambda_k}\tanh\!\left(\!\frac{\beta \Lambda_k}{2}\!\right) \!\!\Bigg\}. \nonumber
\label{TransInvComp}
\end{eqnarray}
Having the magnetic contribution to the Gibbs free energy (\ref{finalMarks3}) the magnetization density follows from the formula: 
\begin{eqnarray}
m = -\frac{1}{2\pi} \int_0^\pi dk \frac{A_k}{\Lambda_k}\tanh\left(\frac{\beta \Lambda_k}{2}\right),
\label{TransMagExact}
\end{eqnarray}
and subsequently, the magnetic susceptibility per spin can be derived according to the relation (\ref{LongSusExact}).

For completeness, let explicitly quote zero-temperature limit of the equation of state for the distortion parameter and the corresponding inverse compressibility:
\begin{align}
\alpha \delta \!+\! p \!&+\! \frac{J_0(1 \!-\! |\kappa| \delta) \!+\! 2h}{4\pi(1 \!-\! |\kappa| \delta)} \, \! E(r) \!+\! \frac{J_0(1 \!-\! |\kappa| \delta) \!-\! 2h}{4\pi(1 \!-\! |\kappa| \delta)} \, \! K(r) \!=\! 0, \label{TransZeroPaper1} \\
\varkappa = \alpha &+ \frac{h}{2\pi(1-|\kappa|\delta)^2} \left[E(r) - K(r)\right] \nonumber \\
\quad & \hspace{-20pt} +\!\! \left[ \! \frac{J_0(1 \!-\! |\kappa| \delta) \!+\! 2h}{4\pi(1 \!-\! |\kappa| \delta)} \, 
\frac{d E(r)}{d z} \!+\! \frac{J_0(1 \!-\! |\kappa| \delta) \!-\! 2h}{4\pi(1 \!-\! |\kappa| \delta)} \, \frac{d K(r)}{d z} \! \right] \! \frac{d r}{d \delta}, \label{TransZeroPaper2}
\end{align}
where $E(r)\equiv \int_0^{\pi/2} dk\sqrt{1-r^2\sin^2k}$ and $K(r)\equiv \int_0^{\pi/2} dk/\sqrt{1-r^2\sin^2k}$ are the complete elliptic integrals of the second and first kind, respectively, with the modulus $r$ defined through
\begin{eqnarray}
r^2=1-\left[\frac{J_0(1-|\kappa|\delta)-2h}{J_0(1-|\kappa|\delta)+2h}\right]^2. 
\label{Zdeff}
\end{eqnarray}
In the equation (\ref{TransZeroPaper2}) we also define their derivatives $d E(r)/dr=\left[E(r) - K(r)\right]/r$, $d K(r)/dr=-K(r)/r + E(r)/\left[r(1-r^2)\right]$ and $dr/d\delta=4J_0 |\kappa| h\left[J_0(1-|\kappa|\delta)-2h\right]/\{r\left[J_0(1-|\kappa|\delta) + 2h\right]^3\}$. The magnetic part of the ground-state energy (per spin) can be derived from the Hamiltonian (\ref{bogoFinal}) by assuming the occupation number $n_{k}=0$ for all $k$:
\begin{eqnarray}
e_m = -\frac{1}{2\pi}\int_{0}^{\pi} dk \Lambda_k,
\label{TransZeroEng}
\end{eqnarray} 
which allows a straightforward derivation of the magnetization per spin:
\begin{align}
m &= -\frac{1}{2\pi} \int_0^\pi dk \frac{A_k}{\Lambda_k}. \label{TransZeroPaper3}
\end{align}
Similarly as at finite temperatures, the magnetic susceptibility per spin can be derived from Eq. (\ref{TransZeroPaper3}) according to the relation (\ref{LongSusExact}).

\begin{figure*}
\begin{center}
\includegraphics[width=0.5\textwidth]{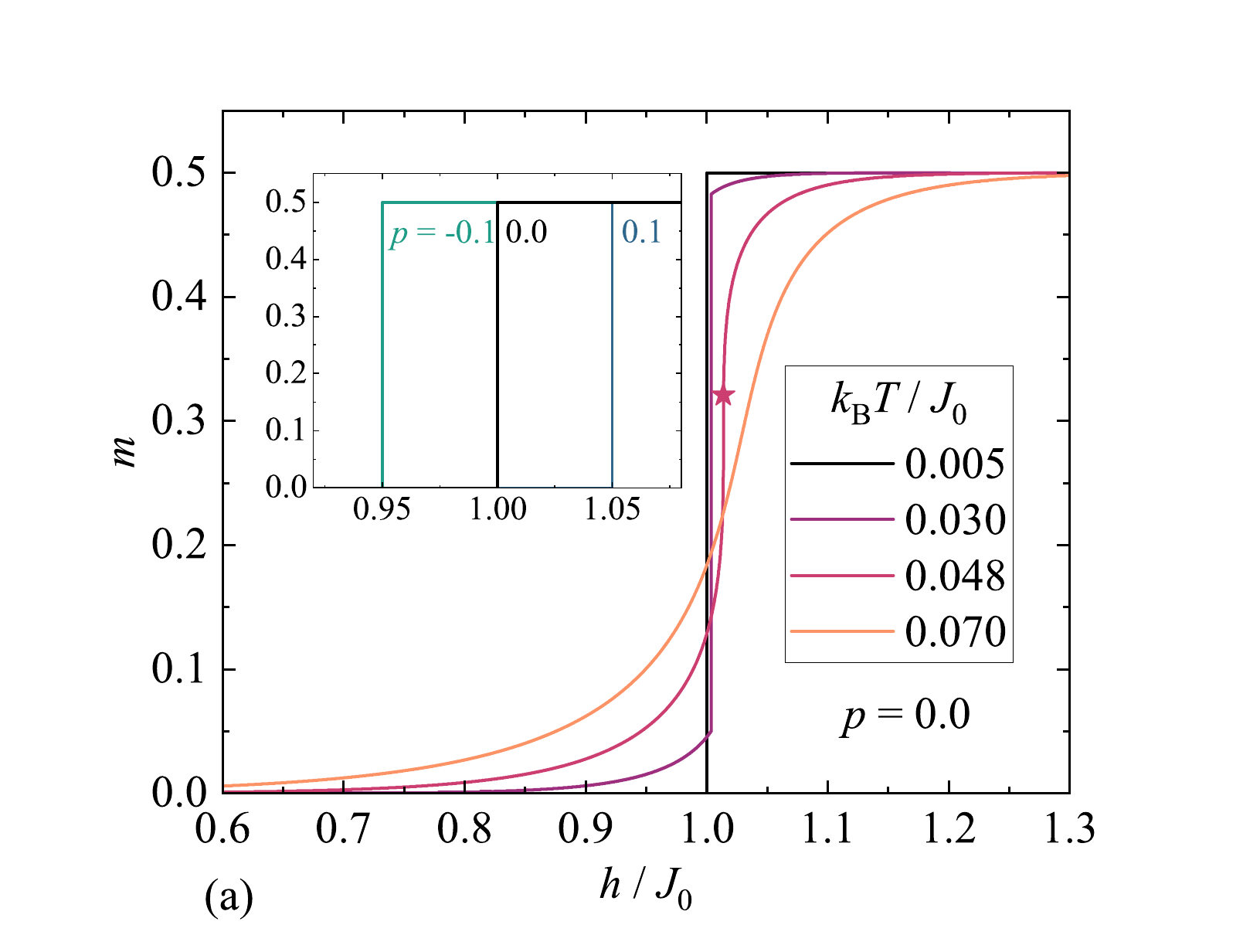}
\hspace*{-1cm}
\includegraphics[width=0.5\textwidth]{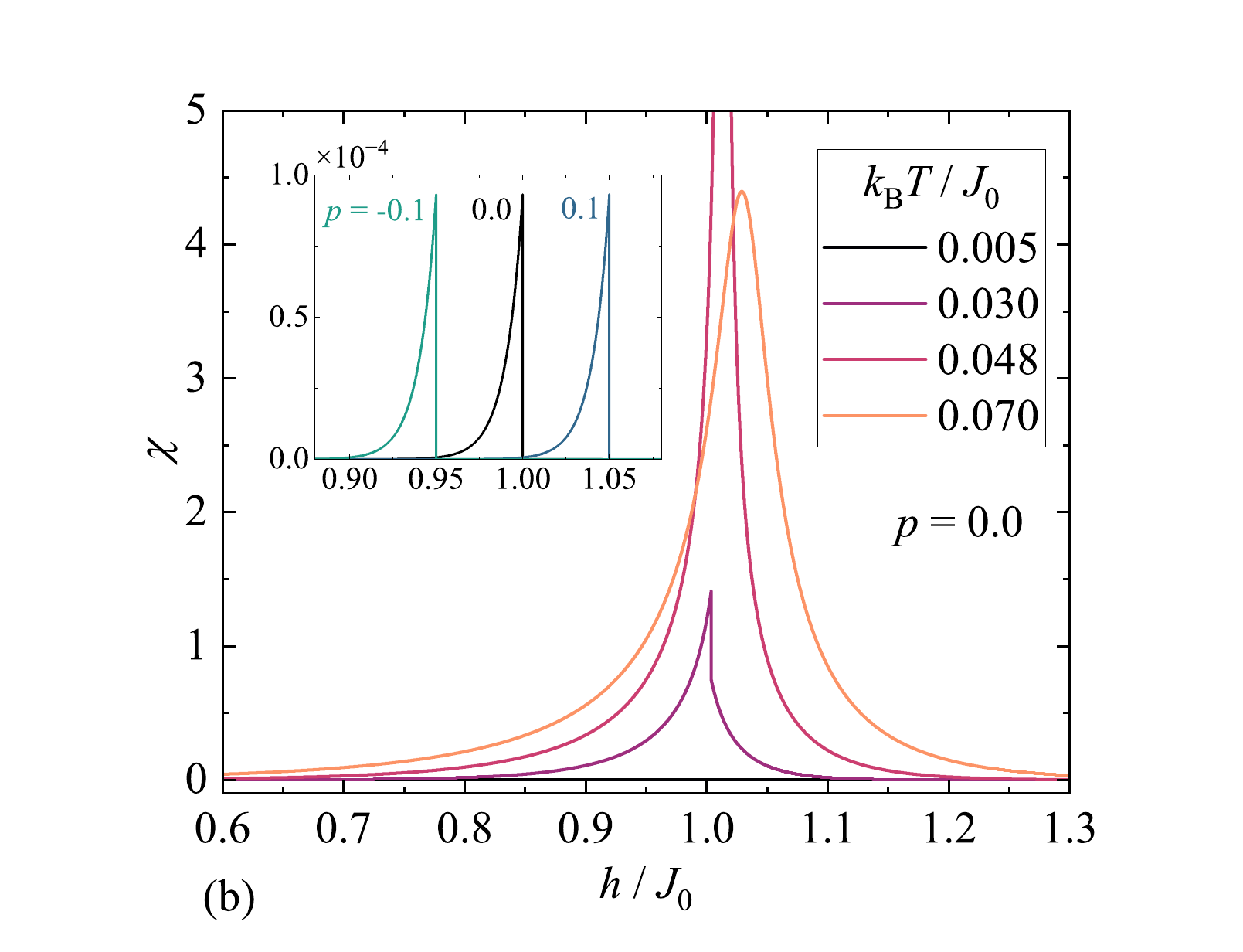}
\includegraphics[width=0.5\textwidth]{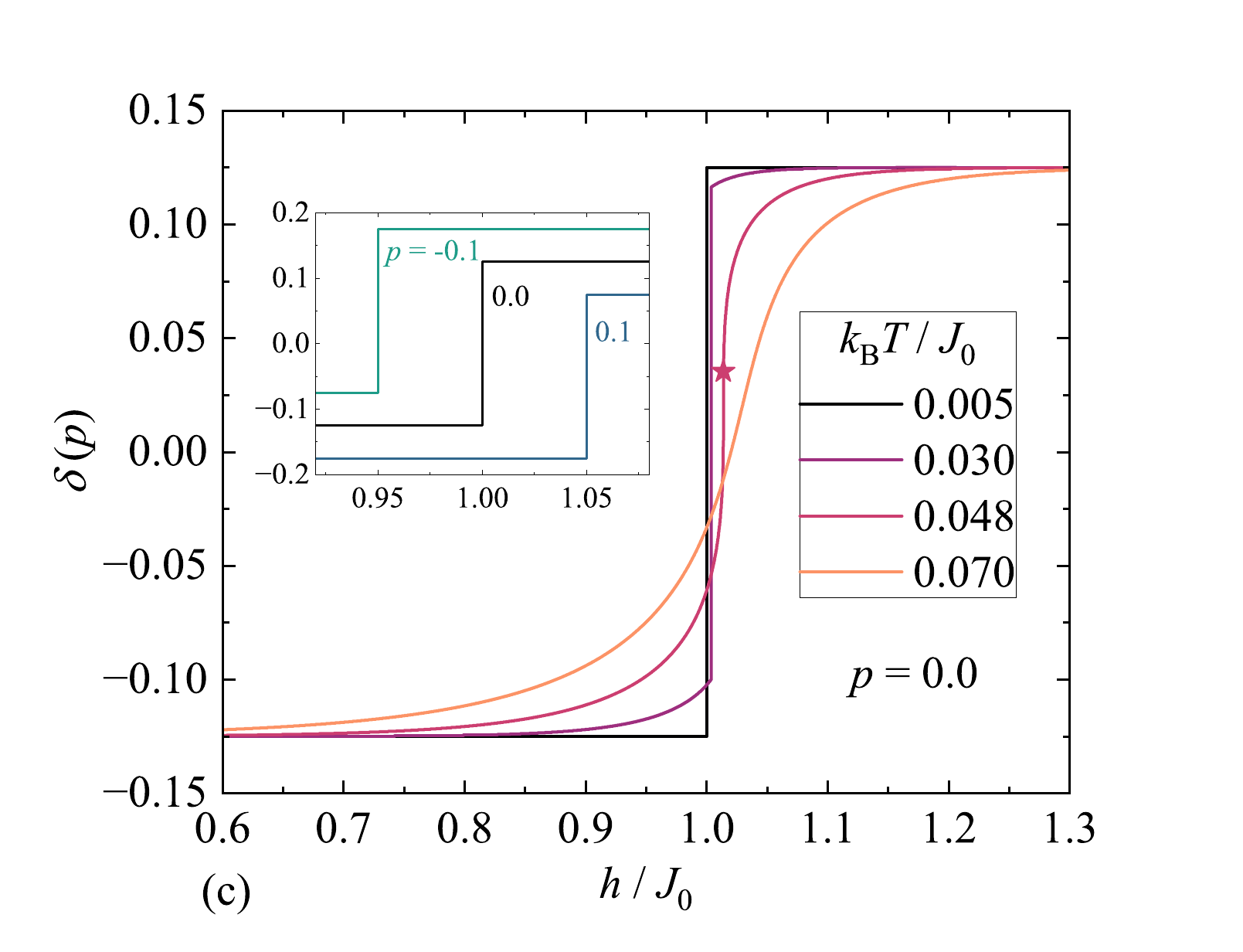}
\hspace*{-1cm}
\includegraphics[width=0.5\textwidth]{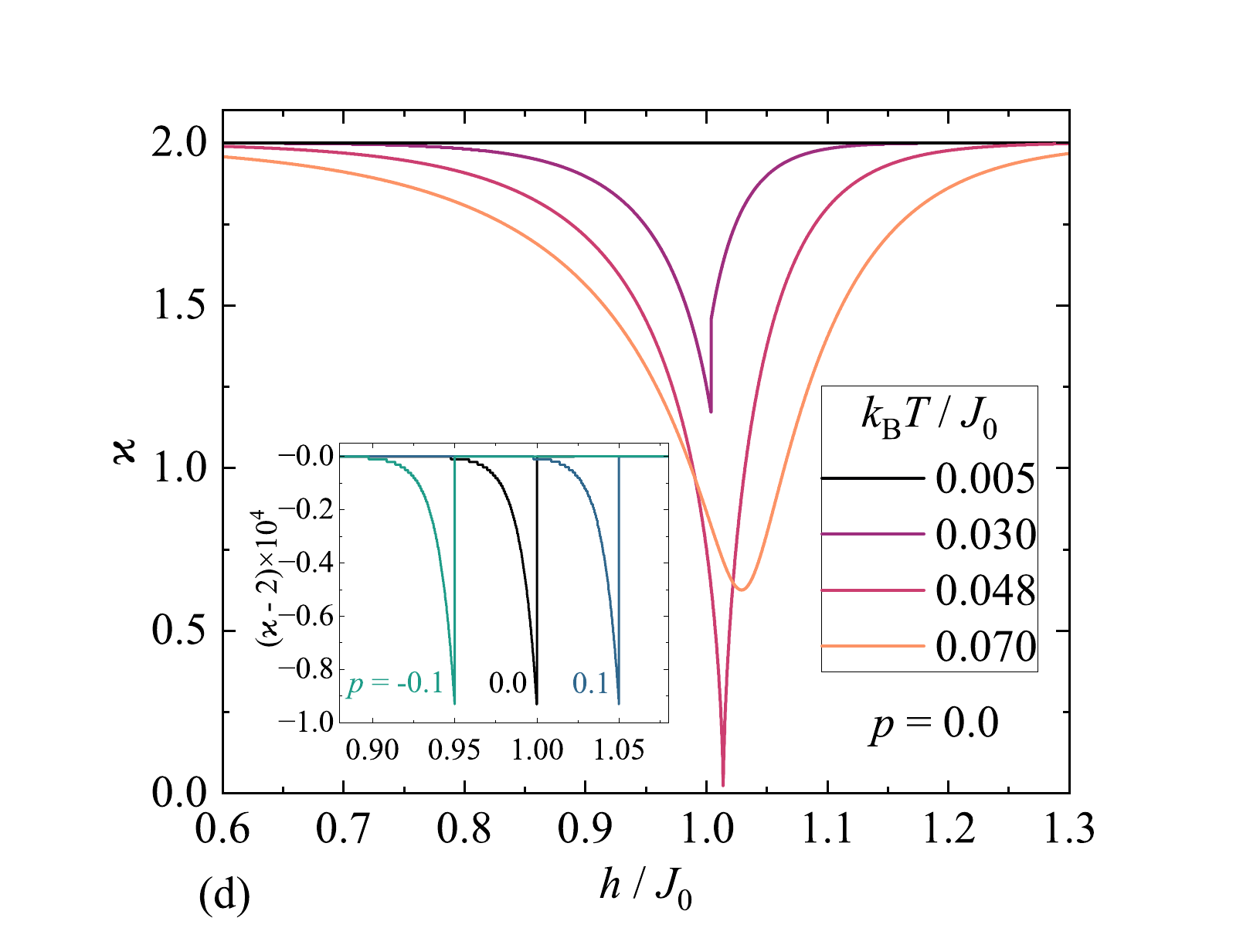}
\end{center}
\vspace{-0.9cm}
\caption{Magnetic-field dependence of (a) the magnetization, (b) the magnetic susceptibility, (c) the distortion parameter, and (d) the inverse compressibility of the deformable spin-1/2 Ising chain in a longitudinal magnetic field at fixed pressure $p = 0$ and four distinct temperatures. For the lowest temperature $k_{\rm B}T/J_0 = 0.005$, the insets display additional results for two specific values of the nonzero pressure $p = \pm 0.1$ for comparison. To better visualize small dip singularities in the inset of Fig. \ref{fig1}(d), the vertical axis represents the deviation  of the inverse compressibility from the reference value $\varkappa = 2$. A star symbol marks the critical point.}
\label{fig1}       
\end{figure*}

\section{Results and discussion}
\label{results}

In this section, we present the most notable results obtained for the deformable spin-1/2 Ising chain subjected to either a longitudinal or a transverse magnetic field. Throughout the rest of this paper, we fix the elastic constant to $\alpha=2$ and the antiferromagnetic coupling constant $J_0 = 1$, which will henceforth serve as the unit of energy when defining the dimensionless magnetic field $h/J_0$ and temperature $k_{\rm B}T/J_0$. To avoid overparametrization, we further set the magnetoelastic coupling constant to $|\kappa| = 1$. Unless stated otherwise, we focus primarily on the case of zero external pressure $p=0$, since positive (negative) pressures only result in quantitative shifts of the elastic and magnetic characteristics without altering the qualitative physical behavior of the system.

\subsection*{A. Longitudinal field}
\label{Rsubsection_A}

To elucidate the magnetic properties of the deformable spin-1/2 Ising chain in a longitudinal magnetic field at zero pressure $p=0$, we present in Figs. \ref{fig1}(a) and \ref{fig1}(b) the magnetic-field dependence of the magnetization and magnetic susceptibility, respectively, for four representative temperatures.  For completeness, additional data corresponding to two specific values of the nonzero pressure $p = \pm 0.1$ are included in the insets for the lowest considered temperature $k_{\rm B}T/J_0=0.005$ to highlight small quantitative shift in the data without fundamentally altering the physical behavior. At the lowest temperature $k_{\rm B}T/J_0=0.005$ and zero pressure, the magnetization remains zero up to the transition field $h/J_0=1$, at which it exhibits a discontinuous jump to full saturation [see Fig. \ref{fig1}(a)]. This abrupt change indicates a discontinuous field-driven thermal phase transition resulting from the magnetoelastic coupling. At an increased temperature $k_{\rm B}T/J_0=0.03$, the main panel of Fig. \ref{fig1}(a) illustrates how this magnetization jump gradually reduces and shifts towards slightly higher magnetic fields. At the critical temperature $k_{\rm B}T_c/J_0\approx 0.048$, the discontinuous jump in magnetization is replaced by a continuous but nonanalytic behavior characterized by an inflection point with a divergent slope. This marks a continuous field-driven thermal phase transition emerging at the critical field $h_c/J_0 \approx 1.014$ as indicated in Fig. \ref{fig1}(a) by the star symbol. For even higher temperatures such as $k_{\rm B}T/J_0=0.07$, the magnetization is continuous throughout the entire magnetic-field range and does not involve any singularity [see Fig. \ref{fig1}(a)]. The change from a discontinuous to a continuous thermal phase transition is more clearly reflected in the magnetic susceptibility shown in Fig. \ref{fig1}(b). For two lowest temperatures $k_{\rm B}T/J_0=0.005$ and $k_{\rm B}T/J_0=0.03$, the susceptibility displays a finite cusp as depicted in the inset and the main panel of Fig. \ref{fig1}(b). These cusp-like anomalies are characteristic of discontinuous field-induced thermal phase transitions, whereas a marked divergence observed at the critical temperature $k_{\rm B}T_c/J_0\approx 0.048$ signals a continuous field-driven thermal phase transition. A further increase in the temperature (e.g., $k_{\rm B}T/J_0=0.07$) results in a finite round maximum in the magnetic susceptibility consistent with crossover behavior rather than a true thermodynamic singularity.

The thermal phase transitions discussed above are also clearly reflected in the magnetic-field dependence of the distortion parameter and the inverse compressibility displayed in Figs. \ref{fig1}(c) and \ref{fig1}(d). As in the previous panels, the main panel contain the data at zero pressure $p=0$ and four representative temperatures, while the insets contain additional results for finite pressures $p=\pm 1$ at the lowest temperature $k_{\rm B}T/J_0=0.005$. Fig. \ref{fig1}(c) demonstrates that, at low enough temperature $k_{\rm B}T/J_0=0.005$, the distortion parameter remains constant as the magnetic field increases until the transition field $h/J_0=1$ is reached, where a sudden jump from the negative value $\delta = -0.125$ to the positive value $\delta = 0.125$ occurs due to a discontinuous field-driven phase transition. Let us highlight that a negative (positive) value of the distortion parameter $\delta<0$ ($\delta>0$) signifies compressive (tensile) strain closely associated with a lattice contraction (elongation). Thus, the magnetic transition is accompanied by an abrupt switching between two distinct lattice regimes. At a higher temperature $k_{\rm B}T/J_0=0.03$, the discontinuity in the distortion parameter becomes smaller and is followed by a gradual increase toward the value $\delta = 0.125$. At the critical temperature $k_{\rm B}T_c/J_0\approx 0.048$, the discontinuity in the distortion parameter entirely vanishes due to a continuous field-driven thermal phase transition, whereas the relevant dependence is a smooth continuous function without any nonanalytic features over the entire field range at even higher temperatures (e.g., $k_{\rm B}T/J_0=0.07$). These findings are fully consistent with the behavior of the inverse compressibility shown in Fig. \ref{fig1}(d). At the two lowest temperatures $k_{\rm B}T/J_0=0.005$ and $0.03$, the inverse compressibility displays a small but sharp dip reflecting the discontinuous field-induced thermal phase transitions. The dip-like anomaly in the inverse compressibility is extremely small at the lowest considered temperature $k_{\rm B}T/J_0=0.005$, therefore, the vertical axis in the inset of Fig. \ref{fig1}(d) is scaled so as to represent the deviation of the inverse compressibility from the reference value $\varkappa = 2$ in order to enhance visibility of the dip-like singularity. As the temperature approaches the critical value $k_{\rm B}T_c/J_0\approx 0.048$, the dip-like anomaly in the inverse compressibility becomes progressively shallower and eventually approaches zero at a continuous field-driven thermal phase transition, whereas the inverse compressibility exhibits a broad rounded minimum characteristic of crossover behavior rather than a dip-like singularity at even higher temperatures (see e.g. $k_{\rm B}T/J_0=0.07$).

\begin{figure*}
\begin{center}
\includegraphics[width=0.5\textwidth]{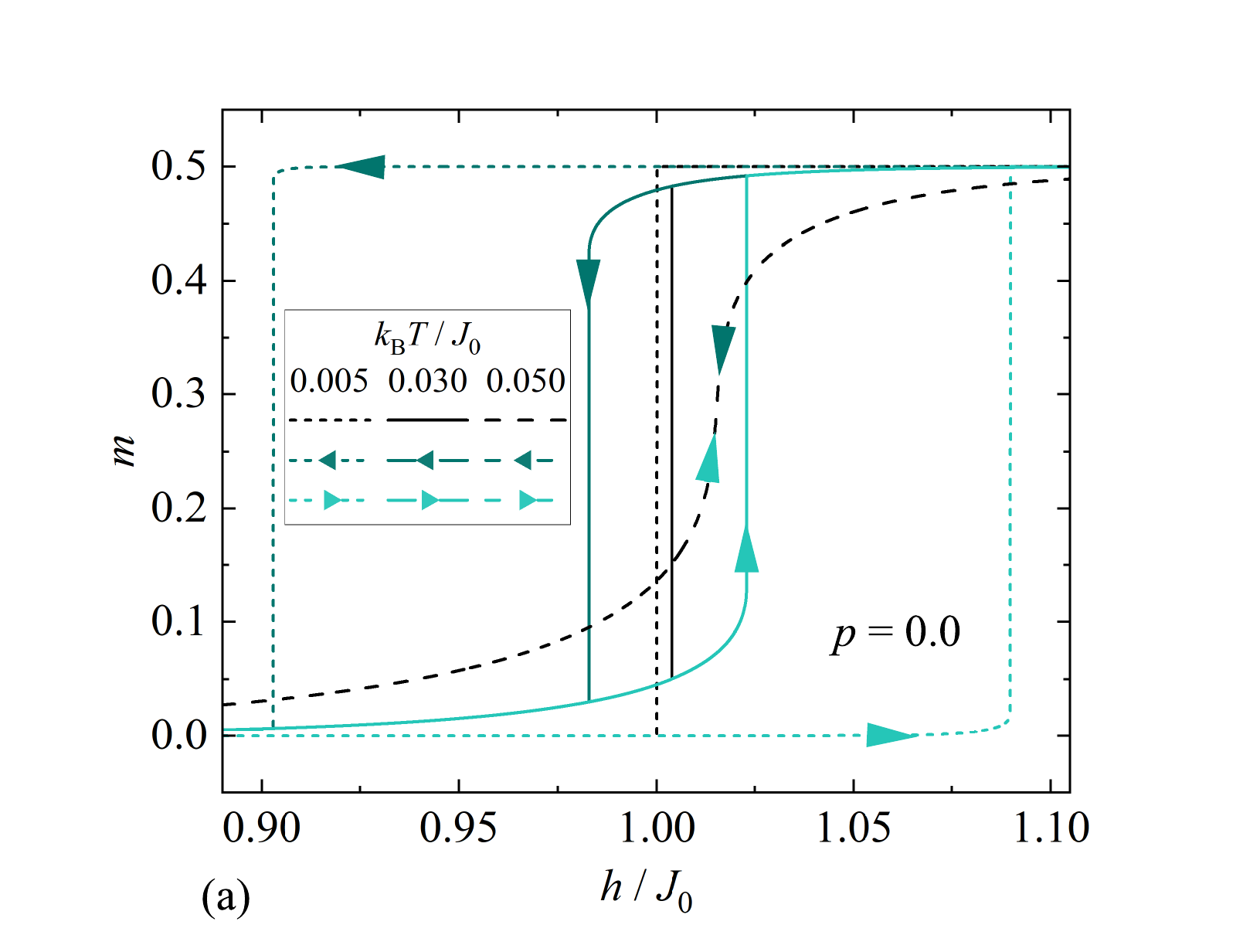}
\hspace*{-1cm}
\includegraphics[width=0.5\textwidth]{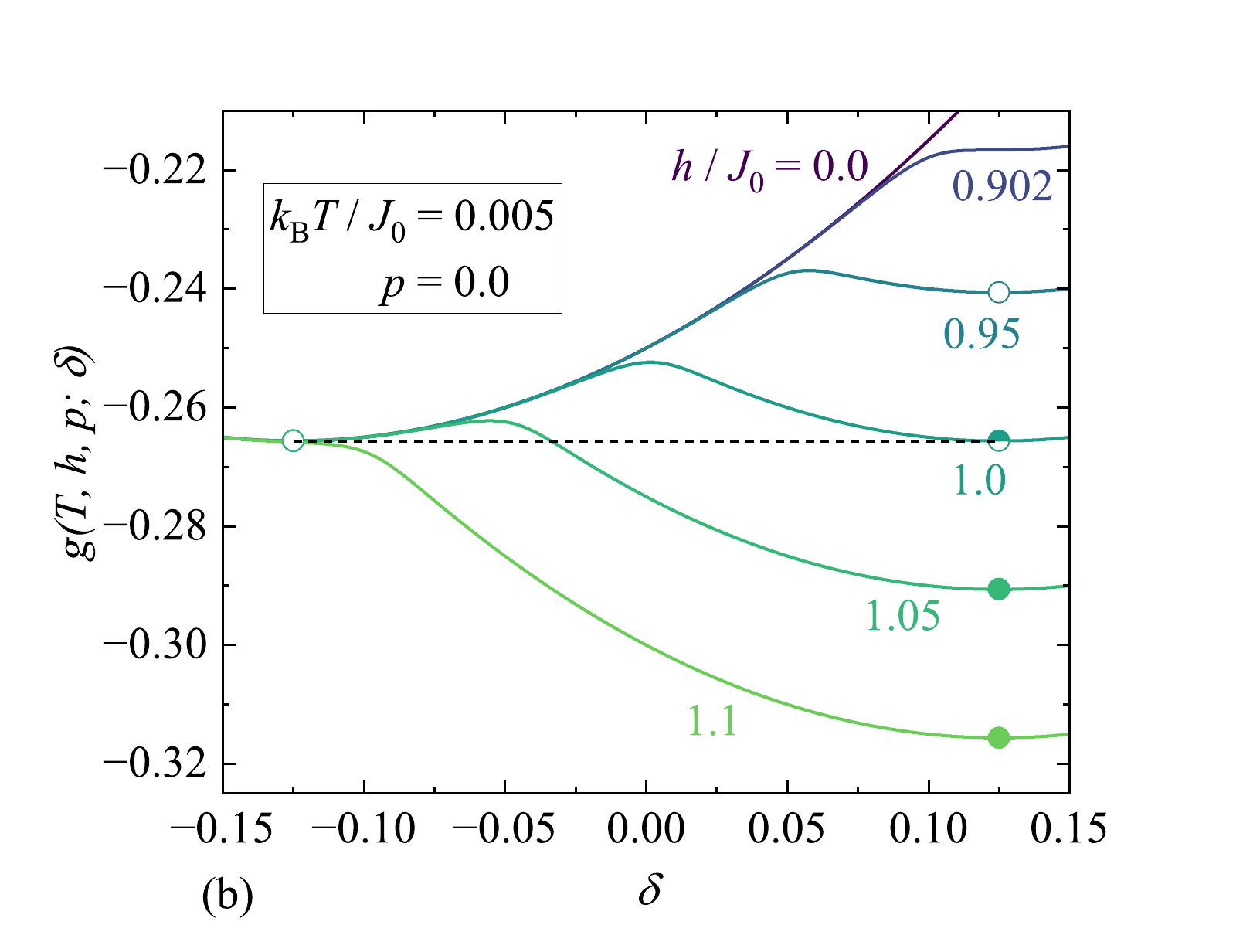}
\includegraphics[width=0.5\textwidth]{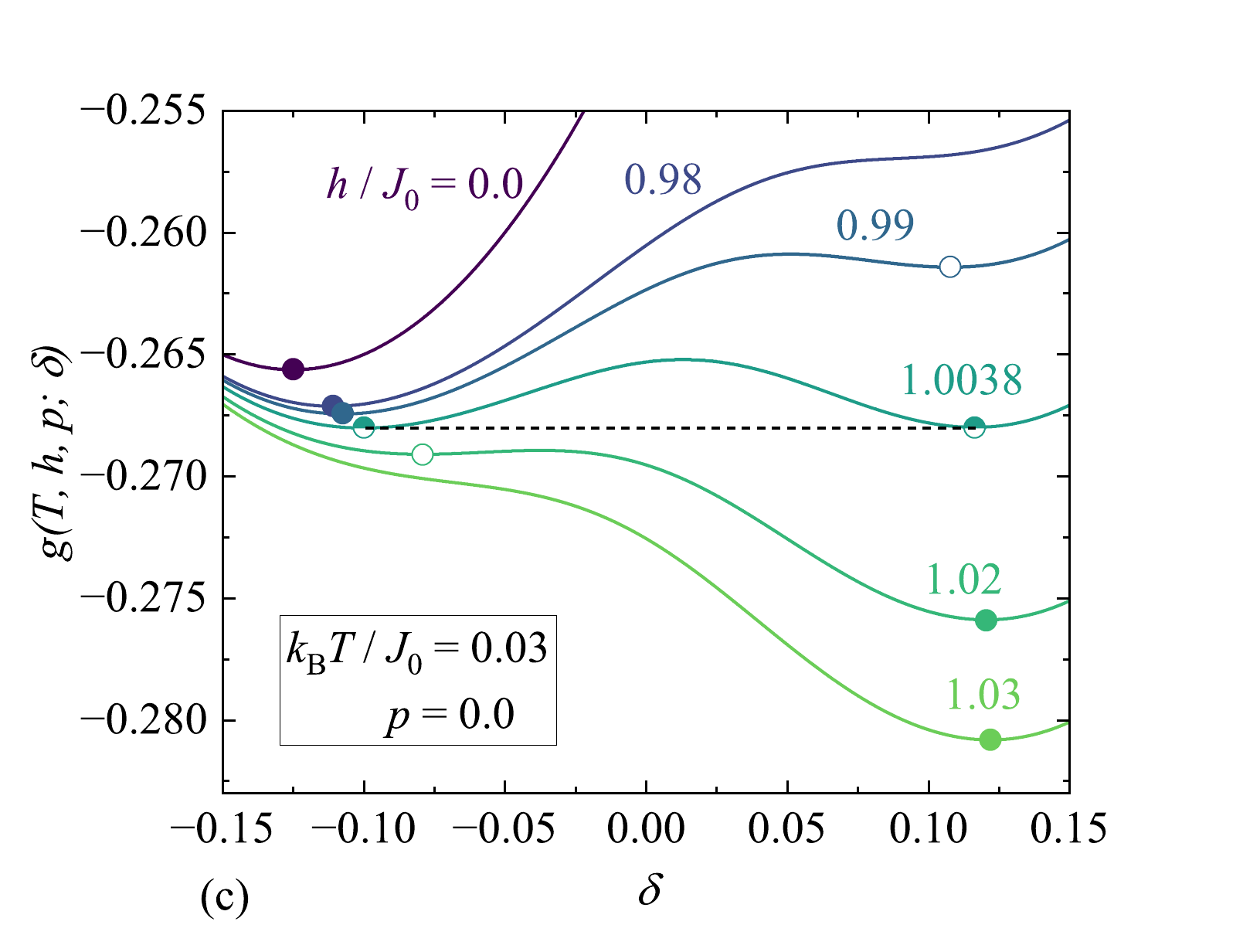}
\hspace*{-1cm}
\includegraphics[width=0.5\textwidth]{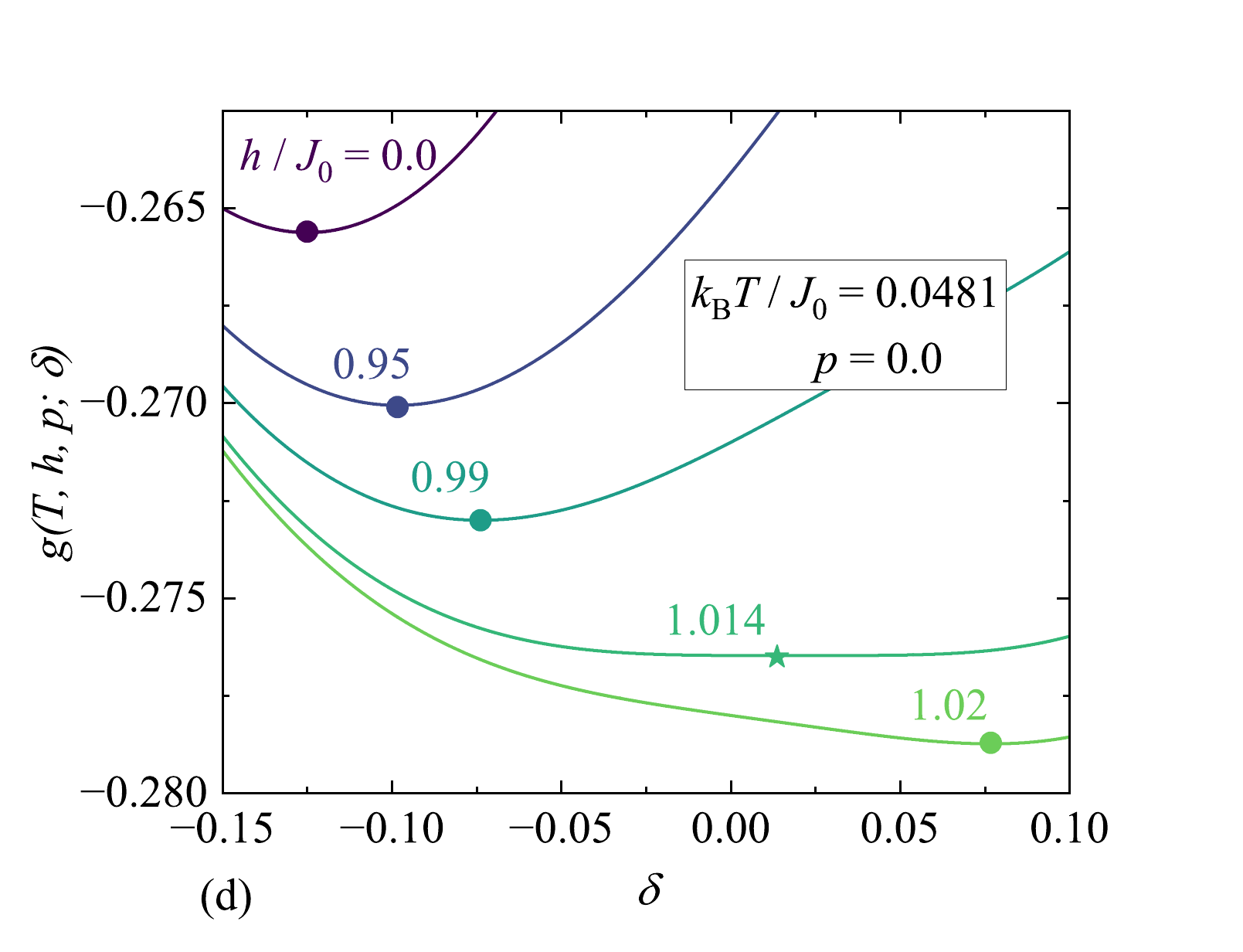}
\end{center}
\vspace{-0.9cm}
\caption{(a) Isothermal field dependence of the magnetization of the deformable spin-1/2 Ising chain in a longitudinal magnetic field at fixed pressure $p=0$ and three distinct temperatures. Left-pointing (right-pointing) arrows indicate decreasing (increasing) field-sweep regime, whereas the curves without any directional arrow correspond to the thermodynamically stable solution obtained from the global minimum of the free energy; (b)-(d) Dependence of the variational Gibbs free energy on the distortion parameter at $p=0$ and three selected temperatures: (b) $k_{\rm B}T/J_0 = 0.005$; (c) $k_{\rm B}T/J_0 = 0.03$; (d) $k_{\rm B}T/J_0 = 0.0481$. Open circles correspond to local minima, filled circles to global minima, and half-filled circles indicate the degeneracy of the two minima. The minimum at the critical point is highlighted with a star symbol.}
\label{fig2}       
\end{figure*}

Discontinuous thermal phase transitions are accompanied by metastable states originating from the existence of multiple local minima in the variational Gibbs free energy, which are consequently manifested in a magnetic hysteresis. It is demonstrated in Fig. \ref{fig2}(a) that the deformable spin-1/2 Ising chain in a longitudinal magnetic field exhibits a hysteresis loop at sufficiently low temperatures $k_{\rm B}T/J_0 \leq 0.0481$ provided that the magnetization is obtained within a quasi-static field-sweep regime following all relevant local minima of the free energy rather than selecting only the global minimum. The magnetization curve for an increasing field sweep (right-pointing arrow in Fig. \ref{fig2}(a)) for the lowest temperature $k_{\rm B}T/J_0 = 0.005$ reveals that the system remains trapped in a metastable state until the upper transition field $h_u/J_0 \approx 1.09$ is reached, where the magnetization discontinuously jumps to its fully saturated value. Conversely, the magnetization jump occurs only at much lower transition field $h_l/J_0 \approx 0.903$ during the (de)magnetization process following a decreasing field sweep regime (left-pointing arrow) resulting in a pronounced hysteresis loop. For comparison, Fig. \ref{fig2}(a) also displays the thermodynamically stable magnetization curve obtained by considering only the global minimum of the variational Gibbs free energy shown by the curve without any directional arrow. To bring insight into the origin of this hysteresis loop, the variational Gibbs free energy is plotted in Fig. \ref{fig2}(b) against the distortion parameter for $k_{\rm B}T/J_0 = 0.005$ and several representative magnetic fields. If the magnetic field lies within the range between the lower and upper transition fields, then, the free energy exhibits both local (open circles) and global (filled circles) minima, whereas only a single minimum is present outside of this field range. Starting from zero field or from fields well above saturation, the deformable Ising chain in the longitudinal field initially occupies the global minimum of the free energy. This global minimum, however, transforms into a local one at $h/J_0 = 1$ as indicated by a thin broken horizontal line connecting two half-filled circles in Fig. \ref{fig2}(b) as the two minima become degenerate at this field strength. This behavior thus provides a compelling evidence for the observed magnetic hysteresis in Fig. \ref{fig2}(a) at $k_{\rm B}T/J_0 = 0.005$. As the temperature increases to $k_{\rm B}T/J_0 = 0.03$, the width of the hysteresis loop decreases as illustrated in Fig. \ref{fig2}(a). This finding is consistent with the shrinking field interval over which two distinct minima coexist as illustrated in Fig. \ref{fig2}(c). At the critical temperature $k_{\rm B}T_c/J_0 \approx 0.0481$ and above (e.g. $k_{\rm B}T/J_0 = 0.05$), the hysteresis loop disappears entirely [see Fig. \ref{fig2}(a)]. This observation provides further compelling evidence for the change from a discontinuous to a continuous thermal phase transition. Indeed, the variational Gibbs free energy shown in Fig. \ref{fig2}(d) at the critical temperature $k_{\rm B}T_c/J_0 \approx 0.0481$ exhibits only a single minimum over the entire field range, which means that the magnetization curve always corresponds to the thermodynamically stable solution independently of the direction of the field sweep. 

\begin{figure*}
\begin{center}
\includegraphics[width=0.5\textwidth]{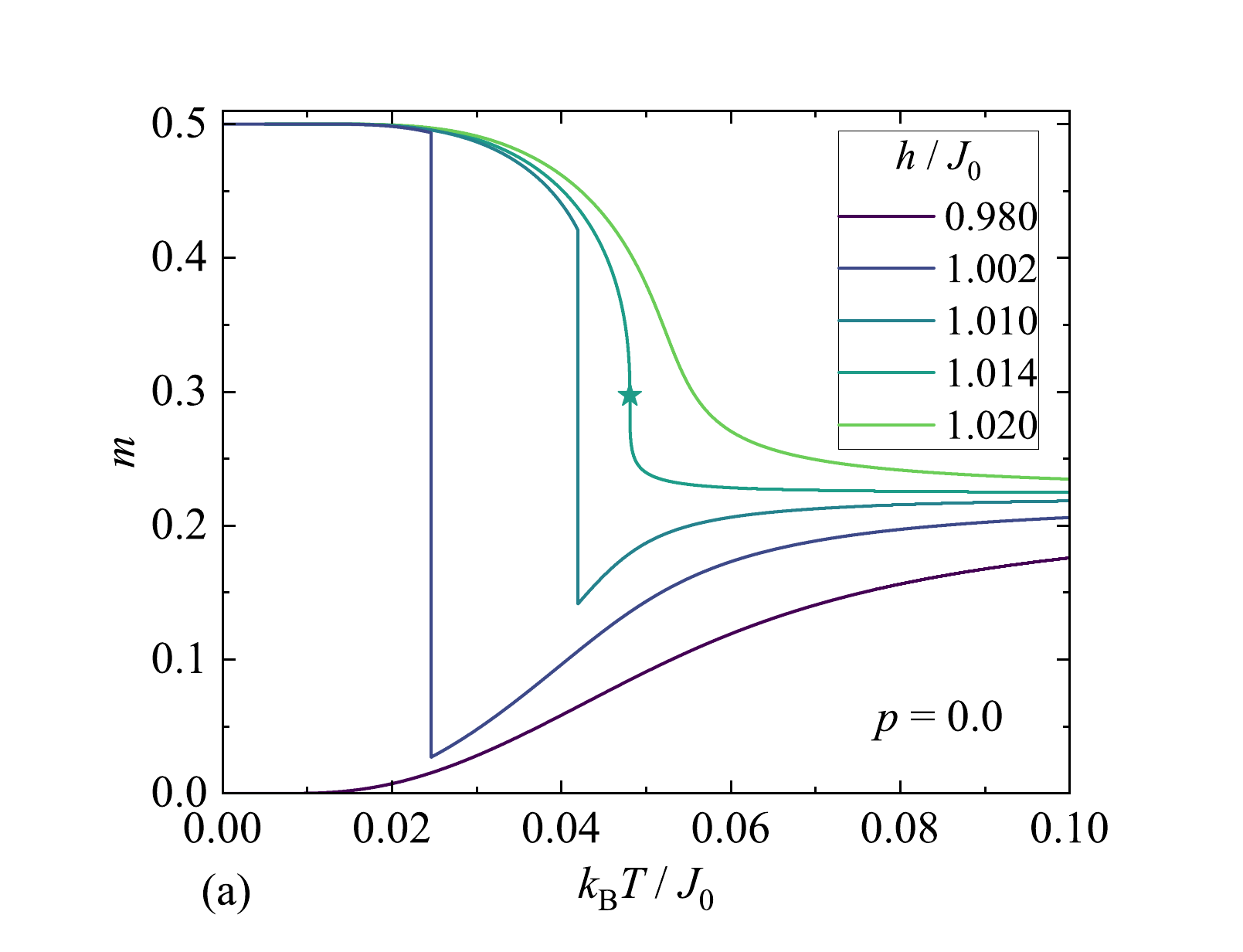}
\hspace*{-1cm}
\includegraphics[width=0.5\textwidth]{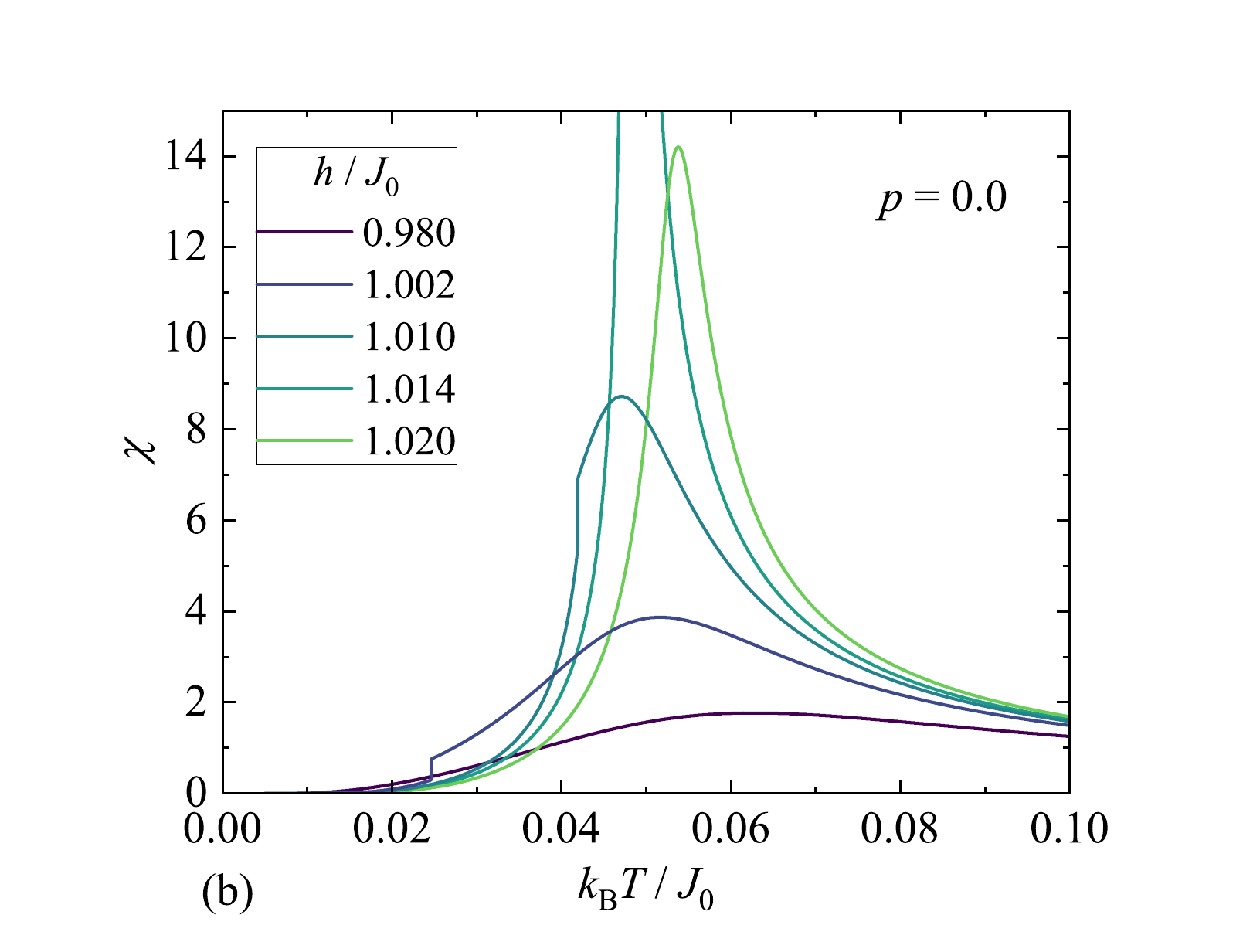}
\includegraphics[width=0.5\textwidth]{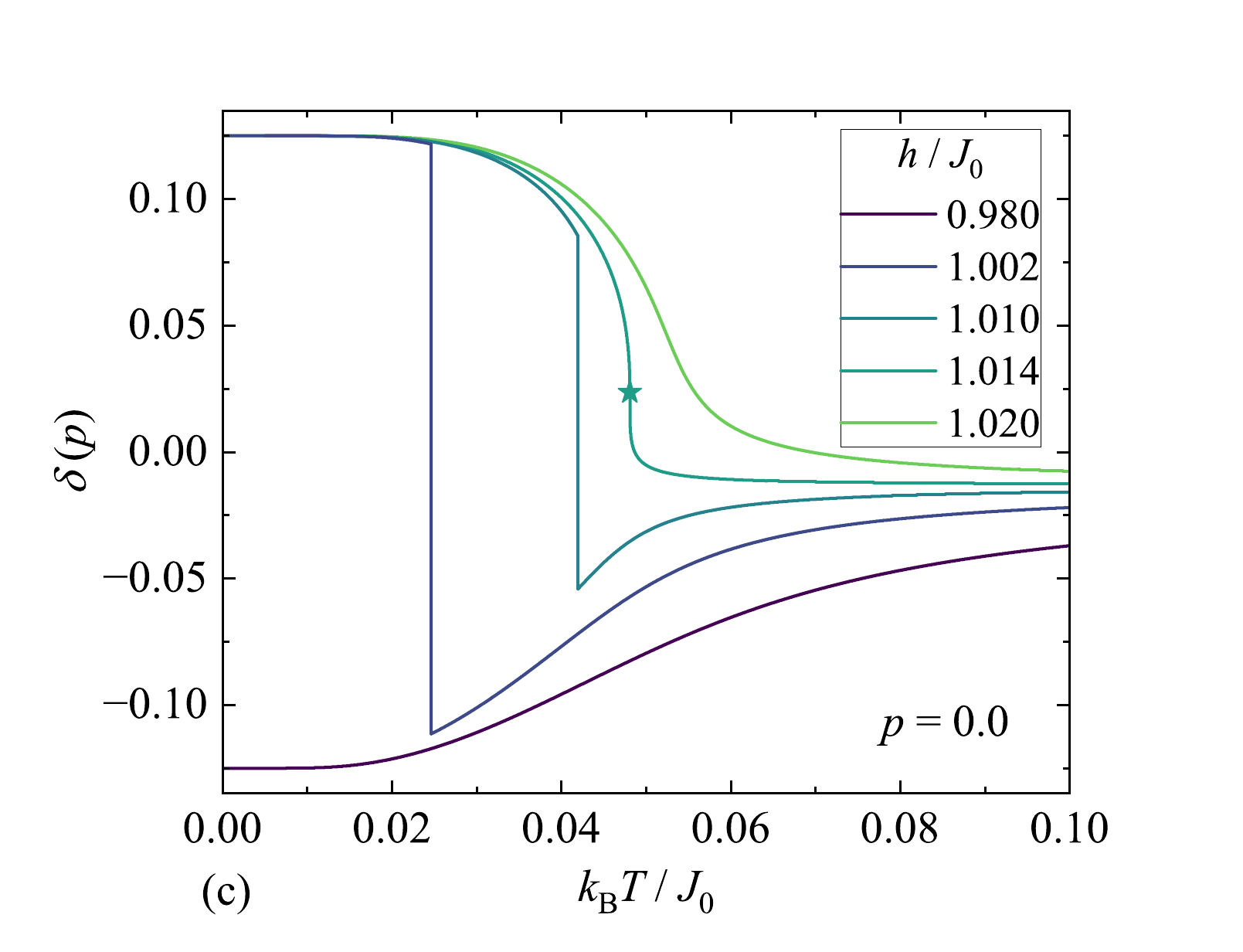}
\hspace*{-1cm}
\includegraphics[width=0.5\textwidth]{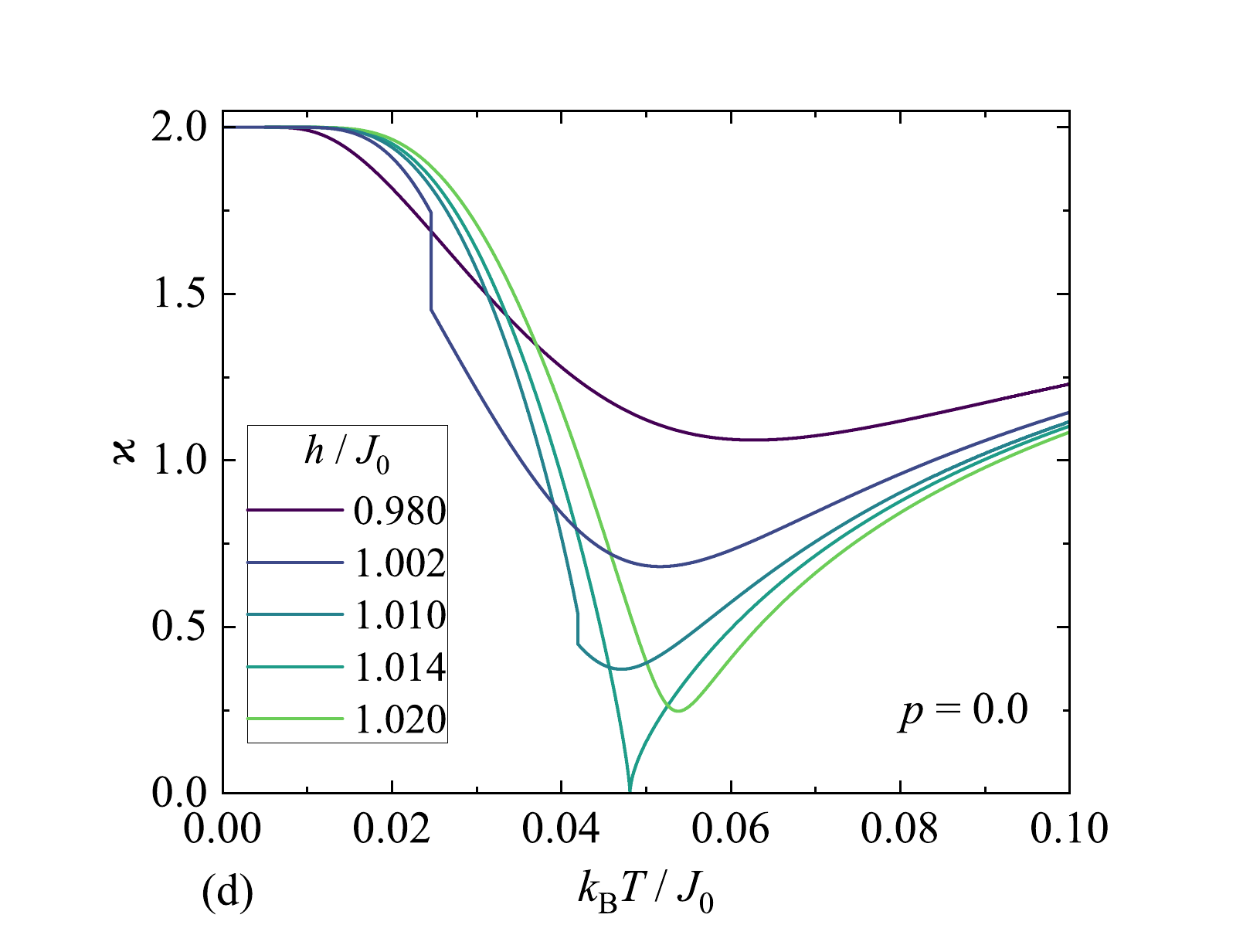}
\end{center}
\vspace{-0.9cm}
\caption{Temperature dependence of (a) the magnetization, (b) the magnetic susceptibility, (c) the distortion parameter, and (d) the inverse compressibility of the deformable spin-1/2 Ising chain at fixed pressure $p=0$ and five selected strengths of the longitudinal magnetic field. A star symbol denotes the critical point.}
\label{fig3}       
\end{figure*}

To gain deeper insight into the behavior of the magnetization and magnetic susceptibility of the deformable spin-1/2 Ising chain in a longitudinal magnetic field, we examine their temperature dependence for several representative field values at fixed pressure $p=0$ as shown in Figs. \ref{fig3}(a) and \ref{fig3}(b), respectively. For the lowest field considered $h/J_0=0.98$, the magnetization increases gradually with temperature without exhibiting any singularity. In contrast, the magnetization curve in Fig. \ref{fig3}(a) displays a discontinuous jump at the transition temperature $k_{\rm B}T/J_0 \approx 0.025$ at the slightly higher field $h/J_0=1.002$ indicating a discontinuous thermal phase transition. As the magnetic field increases further to the value $h/J_0=1.01$, the magnitude of this discontinuity decreases and shifts toward higher temperatures. The discontinuity completely vanishes at the critical field $h_c/J_0 \approx 1.014$ marked by the star symbol in Fig. \ref{fig3}(a), where a continuous phase transition emerges at the critical temperature $k_{\rm B}T_c/J_0 \approx 0.0481$. Beyond the critical field (e.g. $h/J_0=1.02$), the magnetization evolves smoothly without any singularities over the entire temperature range. In agreement with these findings, the magnetic susceptibility shown in Fig. \ref{fig3}(b) exhibits a continuous round maximum for the lowest field $h/J_0=0.98$ characteristic of crossover behavior,  whereas it displays a discontinuity associated with the discontinuous thermal phase transition for larger fields up to $h/J_0 \leq 1.014$. At the critical field $h_c/J_0 \approx 1.014$, the susceptibility reflects through its divergence the emergence of a continuous thermal phase transition. For field strengths exceeding this critical value (e.g. $h/J_0=1.02$), the susceptibility remains continuous across the entire temperature range and develops only a smooth round maximum.

\begin{figure*}
\begin{center}
\includegraphics[width=0.5\textwidth]{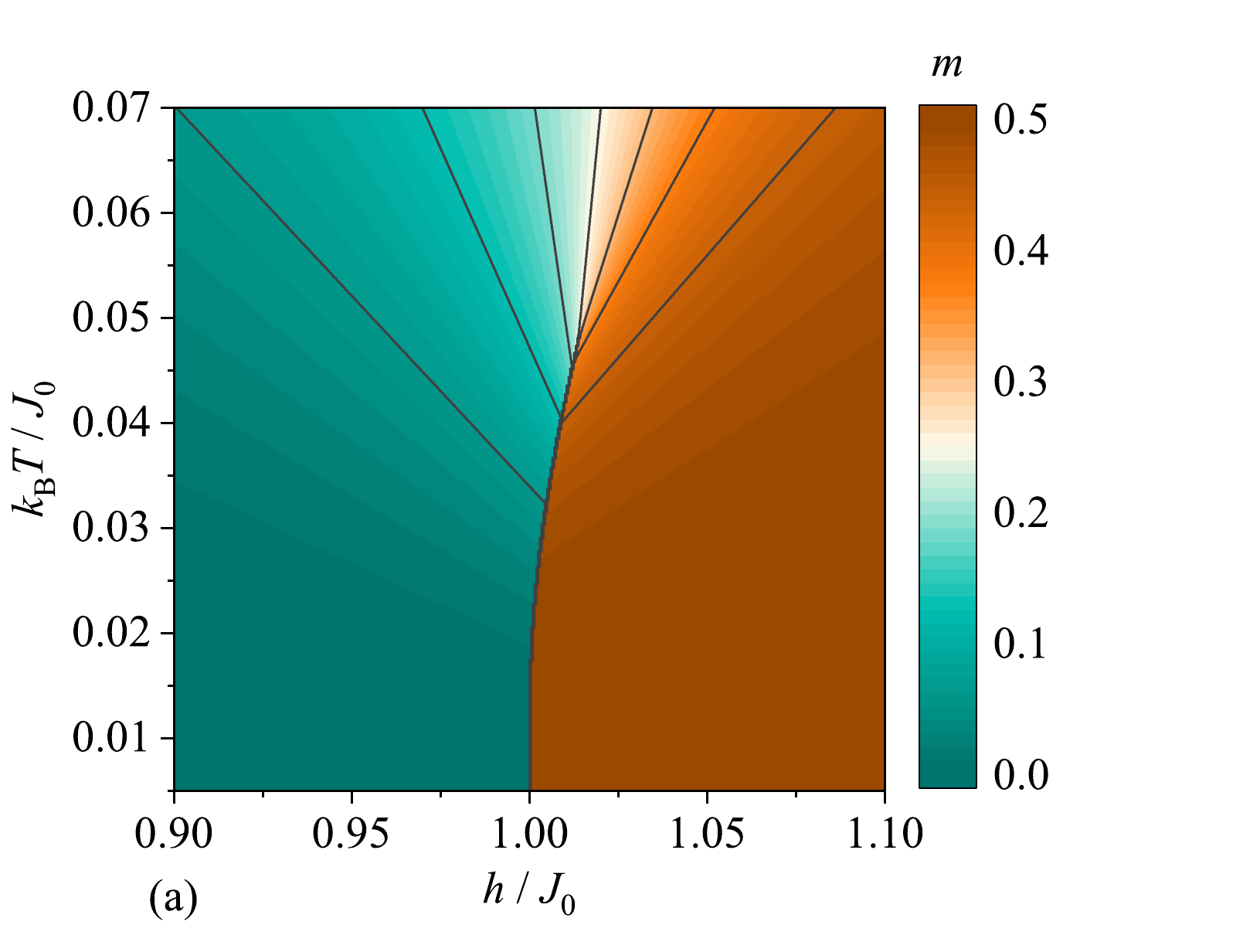}
\hspace*{-1cm}
\includegraphics[width=0.5\textwidth]{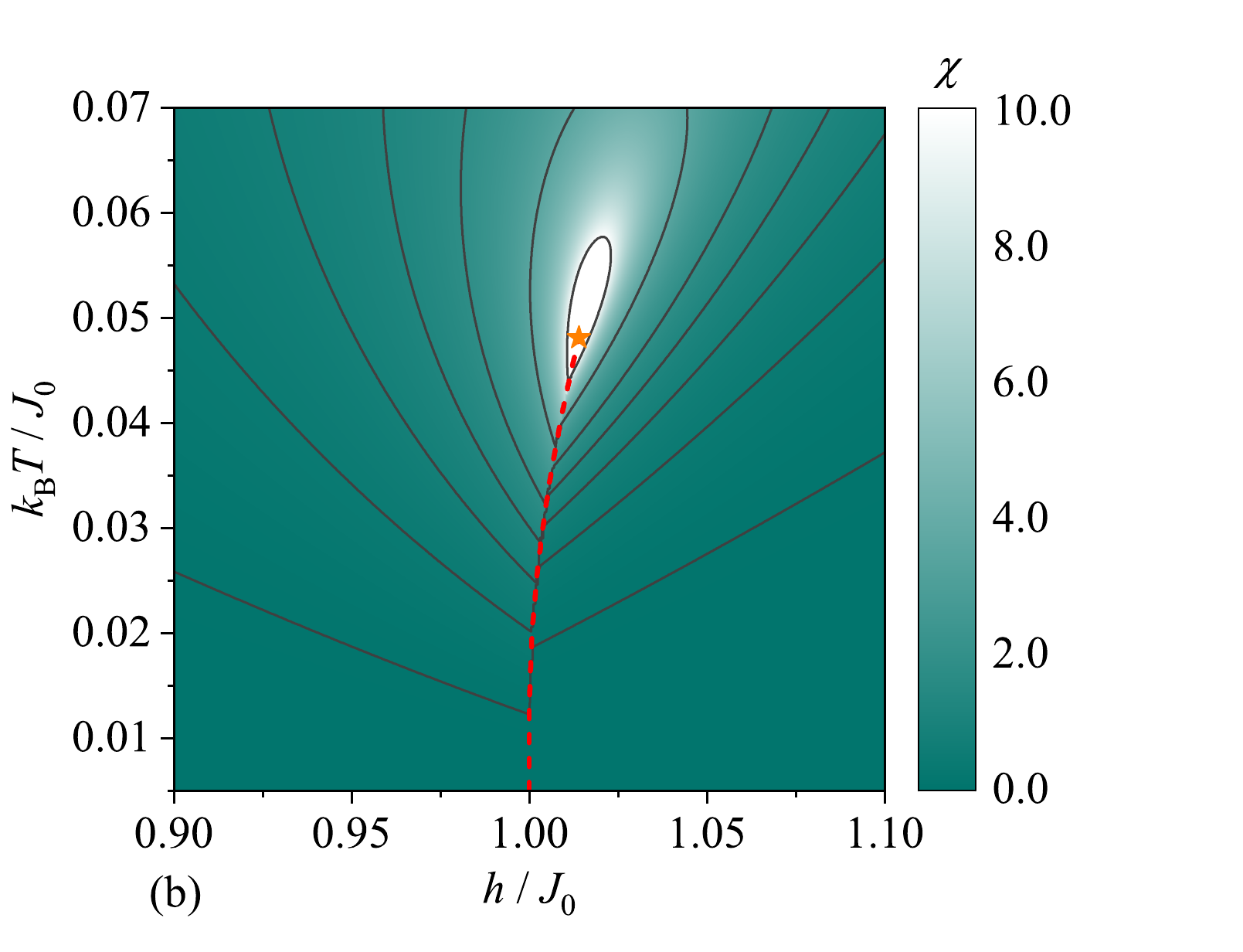}
\includegraphics[width=0.5\textwidth]{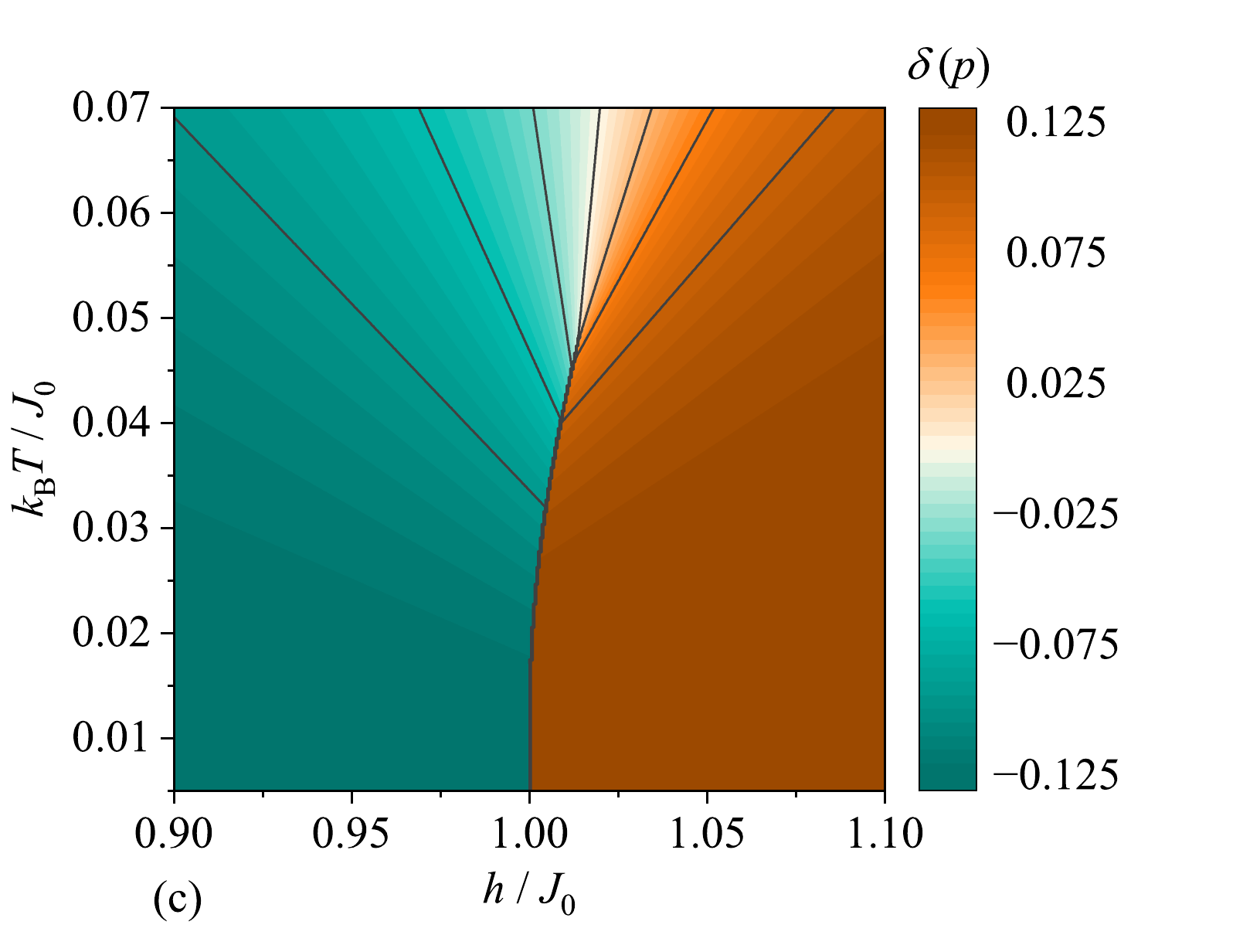}
\hspace*{-1cm}
\includegraphics[width=0.5\textwidth]{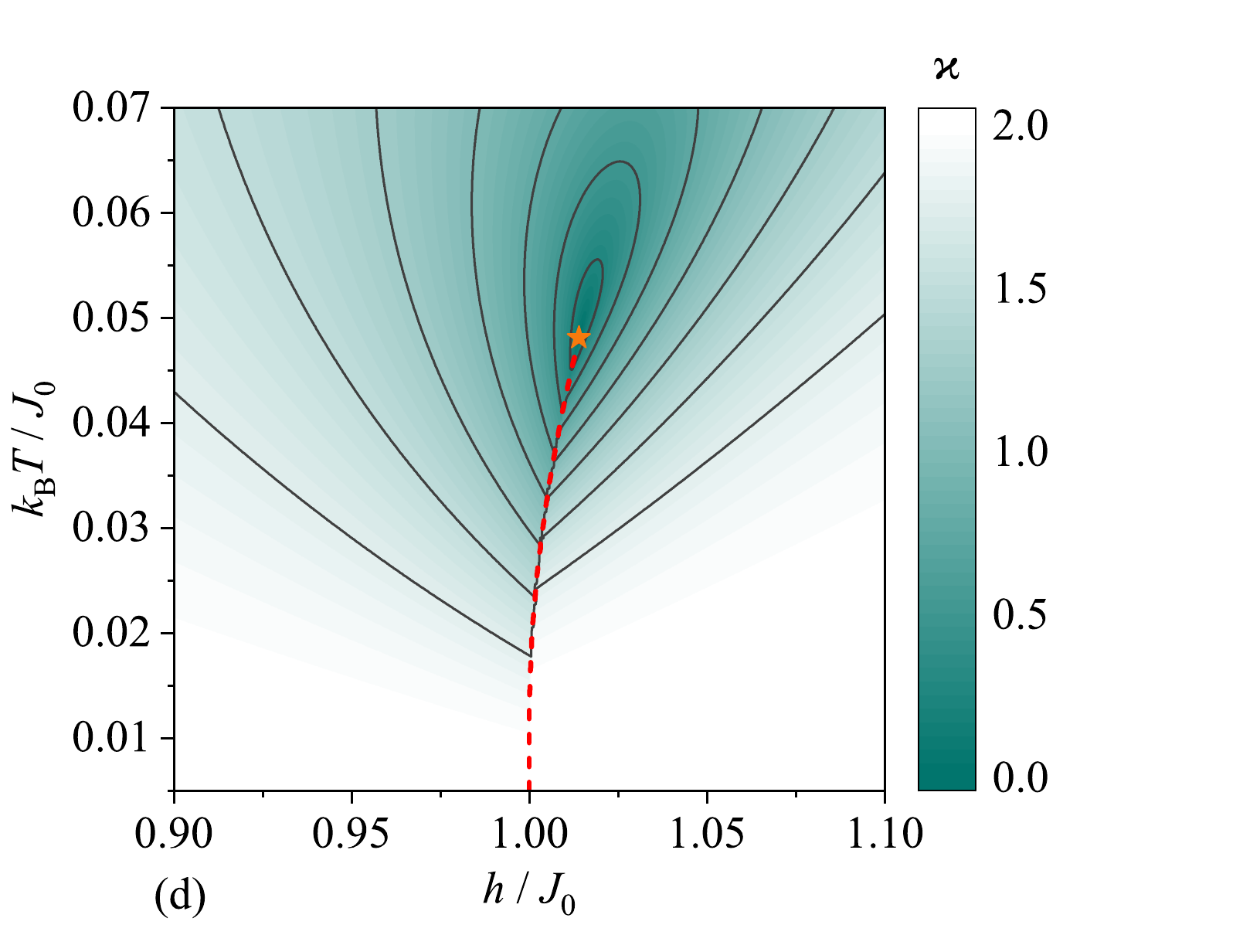}
\end{center}
\vspace{-0.9cm}
\caption{Density plots of (a) the magnetization, (b) the magnetic susceptibility, (c) the distortion parameter, and (d) the inverse compressibility of the deformable spin-1/2 Ising chain in a longitudinal magnetic field shown in the field-temperature plane at fixed pressure $p=0$. In panels (b) and (d), the dashed red curve marks the line of discontinuous thermal phase transitions terminating at the critical point indicated by the orange star symbol.}
\label{fig4}       
\end{figure*}

Figs. \ref{fig3}(c) and \ref{fig3}(d) present the temperature dependencies of the distortion parameter and the inverse compressibility of the deformable spin-1/2 Ising chain at the fixed value of pressure $p=0$ and five distinct values of the longitudinal magnetic field. It can be seen from Fig. \ref{fig3}(a) and (c) that the distortion parameter shows qualitatively similar trends as the magnetization. At the lowest field $h/J_0=0.98$, the distortion parameter increases smoothly with temperature indicating a continuous elongation of the chain without any singular behavior. At higher fields $h/J_0 = 1.002$ and $1.01$, the temperature variations of the distortion parameter shows a discontinuity, which is progressively suppressed and eventually vanishes at the critical field $h_c/J_0 \approx 1.014$ ascribed to a continuous phase transition. For field strengths exceeding this critical value (e.g. $h/J_0=1.02$), the temperature dependence of the distortion parameter becomes smooth again, but now indicating a gradual thermally-assisted suppression of the chain elongation. Fig. \ref{fig3}(d) for the inverse compressibility is a full accordance with the aforementioned behavior of the distortion parameter. At the lowest field considered $h/J_0=0.98$, the inverse compressibility varies smoothly with temperature, whereas it shows an abrupt discontinuous decline for higher fields up to $h/J_0 \leq 1.014$. At the critical field $h_c/J_0 \approx 1.014$, the inverse compressibility completely vanishes as temperature reaches its critical value.  For even larger fields (e.g. $h/J_0=1.02$), the inverse compressibility remains continuous across the entire temperature range and exhibits a rounded minimum.

\begin{figure*}
\begin{center}
\includegraphics[width=0.5\textwidth]{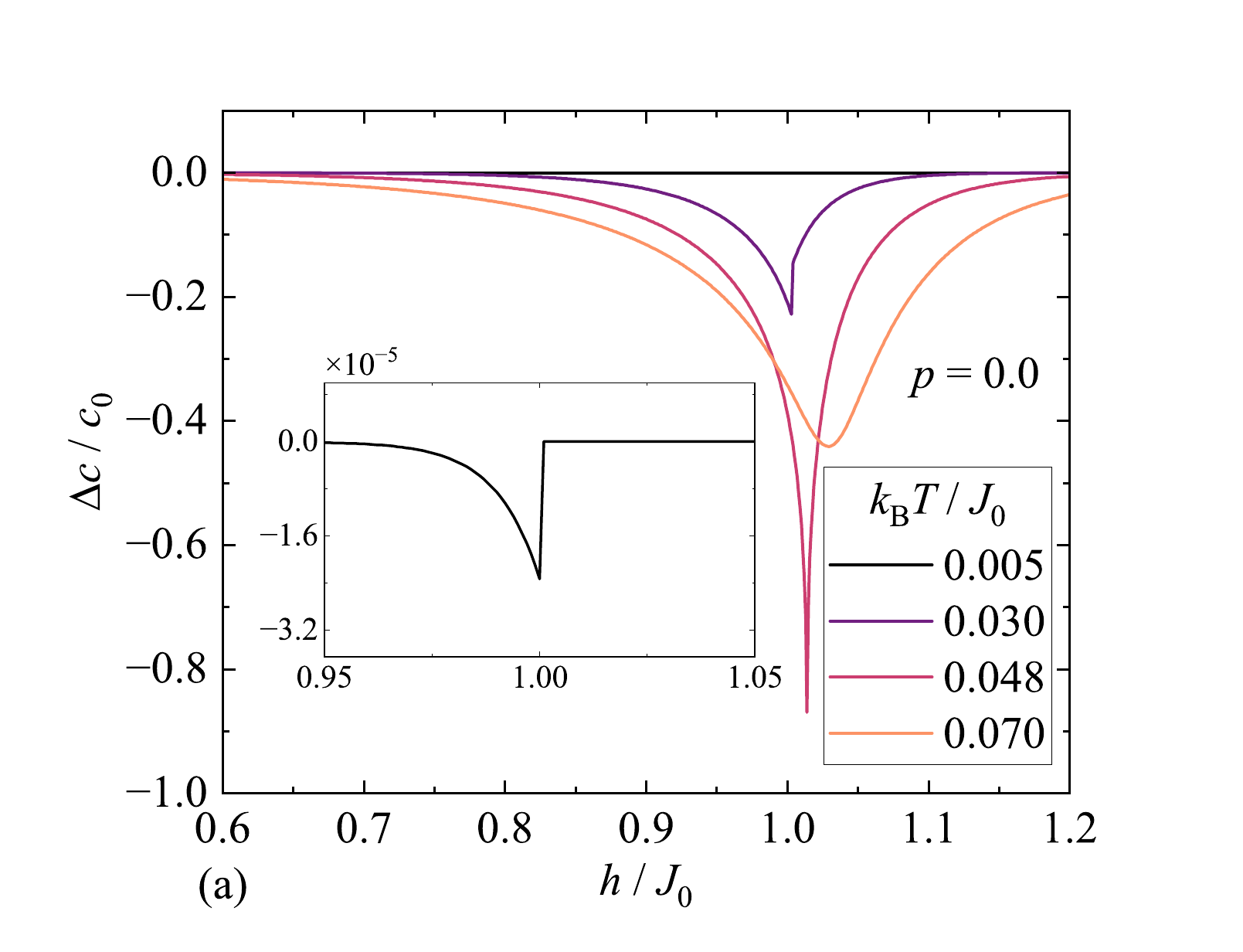}
\hspace*{-1cm}
\includegraphics[width=0.5\textwidth]{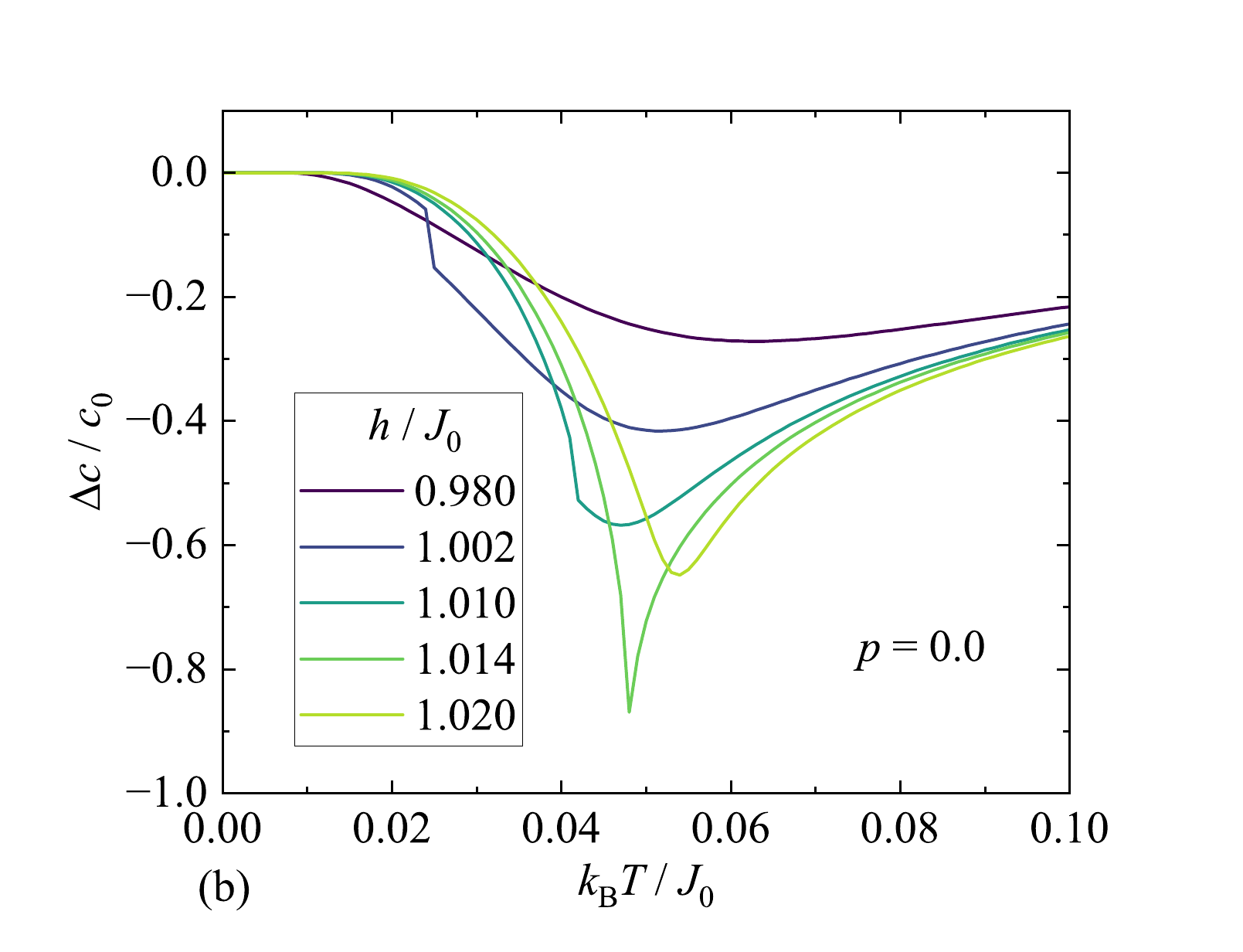}
\end{center}
\vspace{-0.9cm}
\caption{Magnetic-field (a) and temperature (b) dependencies of the relative change of the sound velocity in the deformable spin-1/2 Ising chain under a longitudinal magnetic field at the fixed pressure 
$p = 0$. The inset shows in an enhanced scale a small dip, which appears near the transition field at the lowest considered temperature $k_{\rm B}T/J_0 = 0.005$.}
\label{fig5}       
\end{figure*}

To provide a comprehensive overview of the results presented so far, Figs. \ref{fig4}(a) and \ref{fig4}(b) present density plots of the magnetization and magnetic susceptibility in the field-temperature plane at fixed pressure $p=0$, respectively. Similarly, Figs. \ref{fig4}(c) and \ref{fig4}(d) illustrate the analogous density plots of the distortion parameter and the inverse compressibility. The density plot for the magnetization shown in Fig. \ref{fig4}(a) reveals that all the contour lines branch from a single curve associated with the abrupt change of magnetization, which originates at low temperatures $k_{\rm B}T/J_0 \lesssim 0.02$ from the magnetic field $h/J_0 \approx 1$ and traces the locus of discontinuous phase transitions. The density plot for the magnetic susceptibility shown in Fig. \ref{fig4}(b) demonstrates that this curve indeed represents the line of discontinuous thermal phase transitions (dashed red line), which bends toward higher magnetic fields with increasing temperature. This transition line terminates at the critical point (orange star symbol), which marks a continuous thermal phase transition with the coordinates $h_c/J_0 \approx 1.0140$ and $k_{\rm B}T_c/J_0 \approx 0.0481$ given with a four-digit precision. The density plot for the magnetic susceptibility thus evidences that the thermal phase transitions occur only within a rather narrow field range from $h/J_0 = 1$ up to the critical field $h_c/J_0 \approx 1.014$, whereas the system exhibits smooth crossover behavior without singularities outside this field range. Notably, similar conclusions can be drawn from the density plots for the distortion parameter and the inverse compressibility illustrated in Figs. \ref{fig4}(c) and \ref{fig4}(d), which display analogous signatures of the discontinuous phase transition line and its termination at the critical point.

\begin{figure*}
\begin{center}
\includegraphics[width=0.5\textwidth]{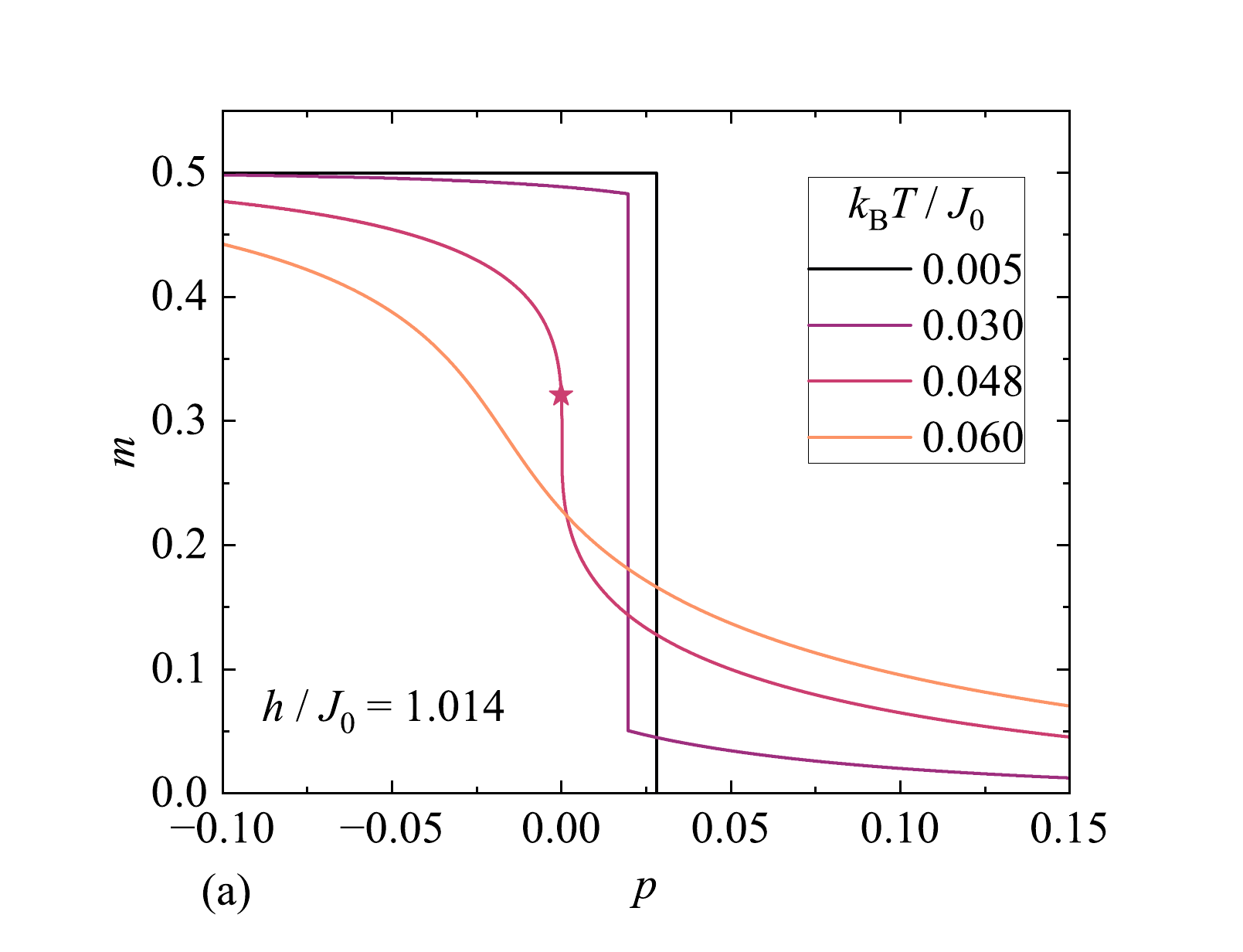}
\hspace*{-1cm}
\includegraphics[width=0.5\textwidth]{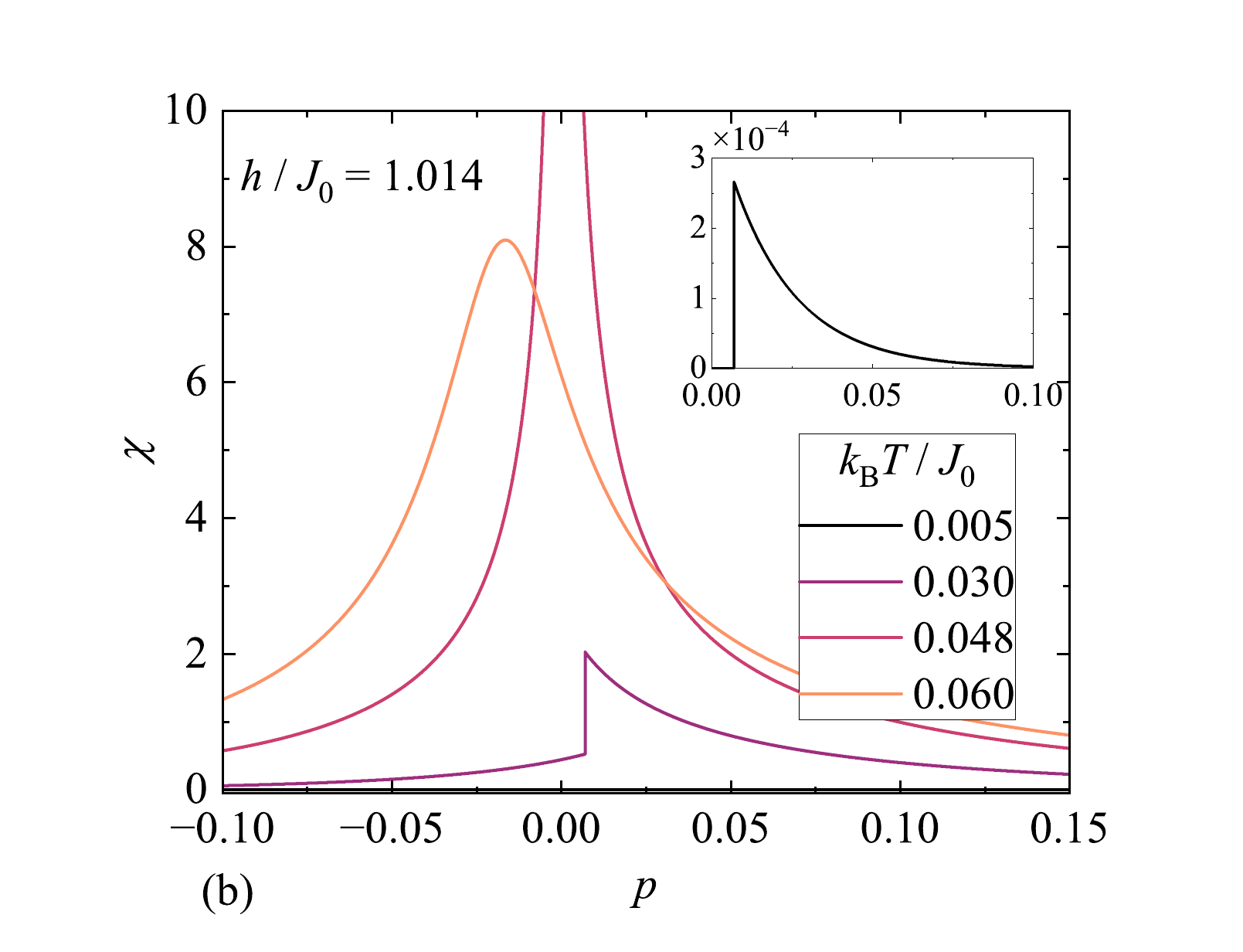}
\includegraphics[width=0.5\textwidth]{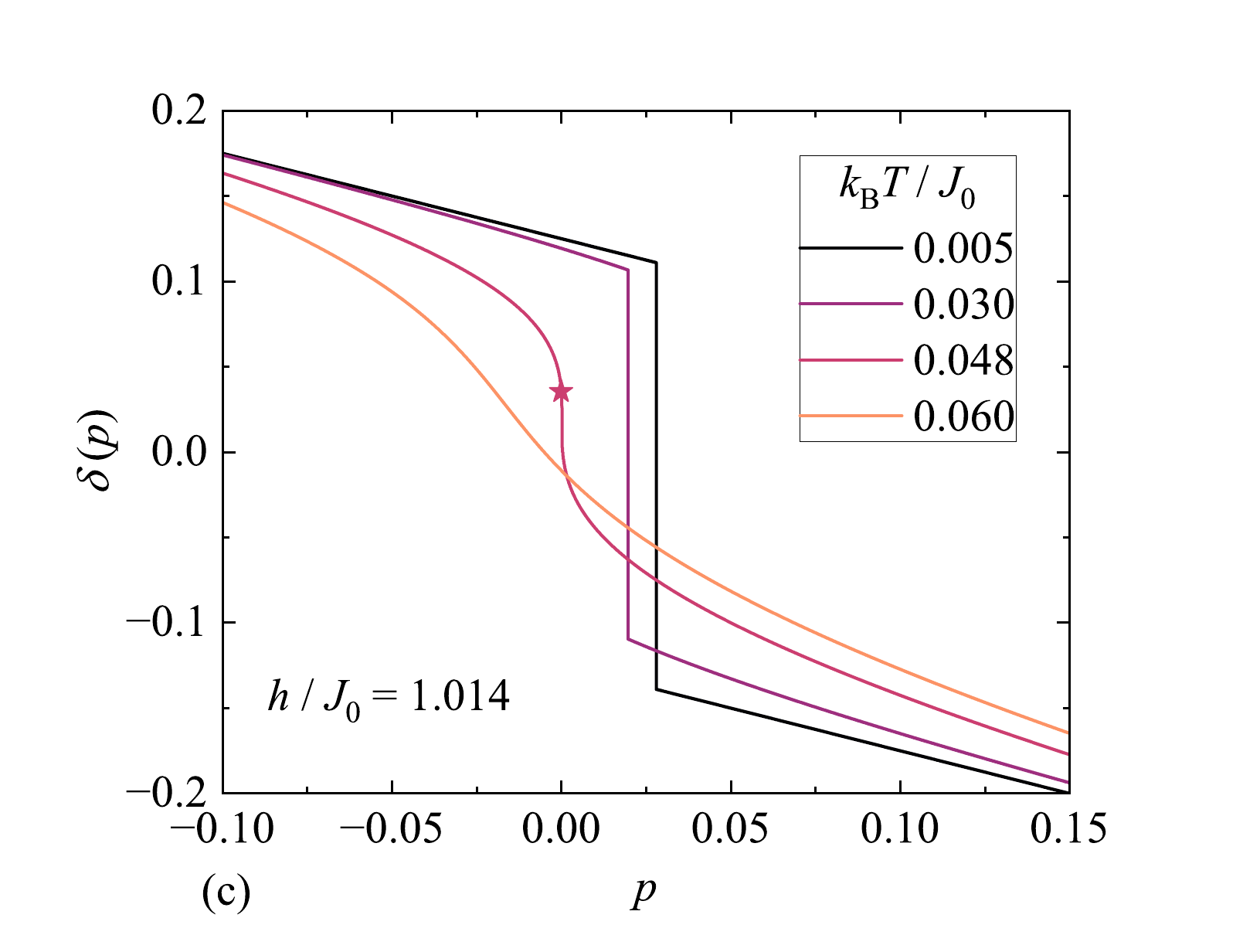}
\hspace*{-1cm}
\includegraphics[width=0.5\textwidth]{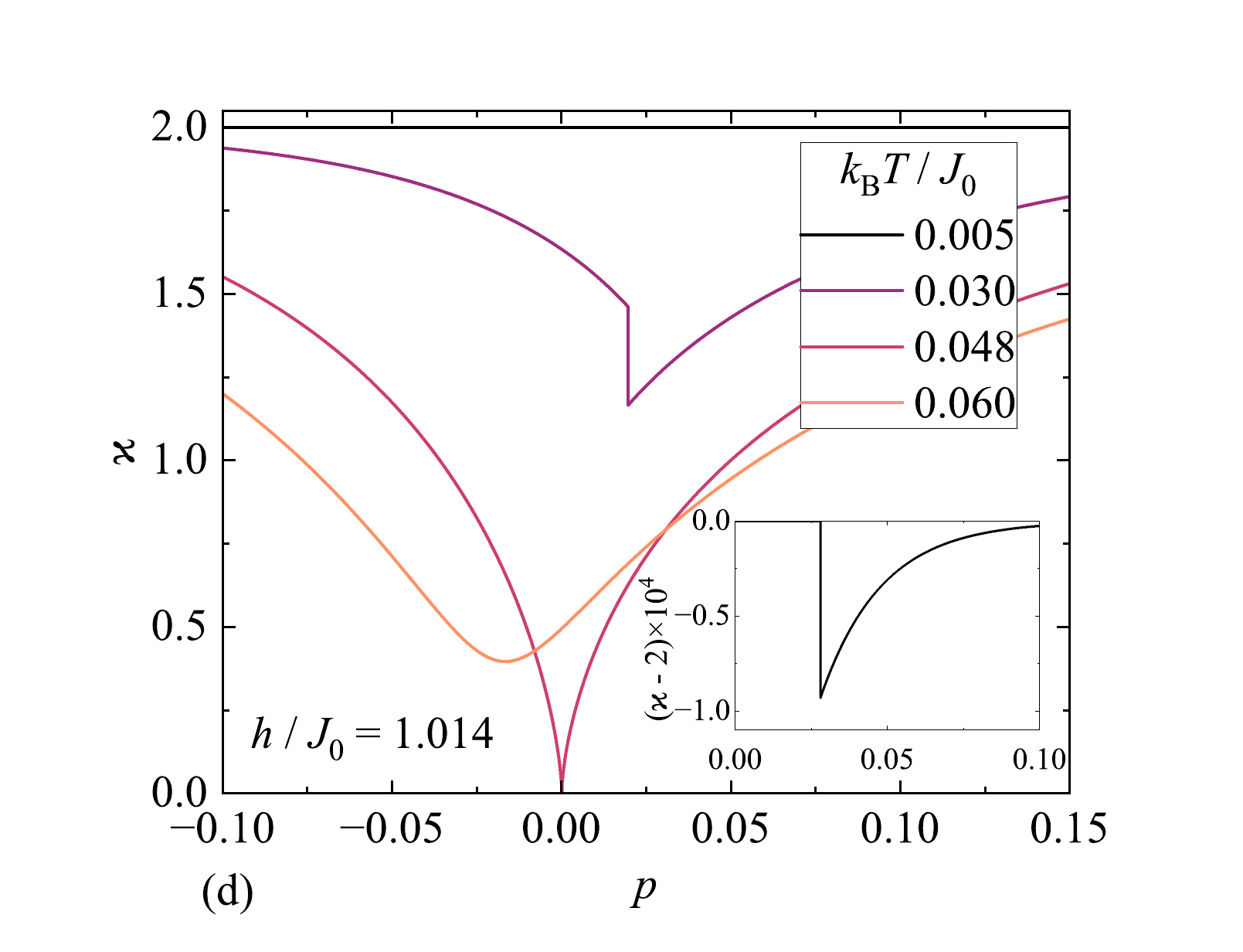}
\end{center}
\vspace{-0.9cm}
\caption{Pressure-induced changes of (a) the magnetization, (b) the magnetic susceptibility, (c) the distortion parameter, and (d) the inverse compressibility of the deformable spin-1/2 Ising chain in a longitudinal magnetic field $h/J_0=1.014$ at four distinct temperatures. In the inset of panel (d), the vertical axis shows the deviation of the inverse compressibility from the reference value $\varkappa = 2$ in order to enhance the visibility of the dip-like singularity. The star symbol denotes the critical point.}
\label{fig6}       
\end{figure*}

Furthermore, Figs. \ref{fig5}(a) and \ref{fig5}(b) present the magnetic-field and temperature dependence of the relative change in the sound velocity calculated according to Eq. (\ref{deltaSound}) for the deformable spin-1/2 Ising chain in a longitudinal magnetic field at fixed pressure $p = 0$. Note that the value $\Delta c / c_0 = -1$ corresponds to a maximal sound attenuation when the sound velocity reaches its global minimum $c = 0$. Both the field and temperature dependencies of the relative change in the sound velocity shown in Fig. \ref{fig5}(a) and (b) closely resemble the corresponding behavior of the inverse compressibility discussed previously [see Figs. \ref{fig1}(d) and \ref{fig3}(d)]. At a discontinuous thermal phase transition, the relative change in the sound velocity displays a discontinuous drop reflected in the dip-like anomaly, whereas the sound velocity gradually converges to its minimum value at a critical point associated with the continuous thermal phase transition. This qualitative correspondence could be expected on the grounds of Eq. (\ref{deltaSound}), which connects the relative change in the sound velocity with a square root of the inverse compressibility that tends to zero at the critical point upon tuning either temperature or magnetic field. It should be pointed out, moreover, that the complete sound attenuation at the critical point requires extremely fine tuning of both temperature and magnetic field to their precise critical values, whereby achieving such conditions experimentally would be highly challenging due to the need for exceptionally small control increments in the vicinity of the critical point.

To complete the overall picture, we demonstrate that a change in pressure itself can also induce a thermal phase transition. Figs. \ref{fig6}(a) and \ref{fig6}(b) display the pressure dependence of the magnetization and magnetic susceptibility of the deformable spin-1/2 Ising chain in a longitudinal magnetic field $h/J_0=1.014$ at four distinct temperatures. The particular value of the magnetic field was selected to coincide (with a four-digit precision) to the critical field at zero pressure $p=0$. At the lowest temperature $k_{\rm B}T/J_0 = 0.005$, the magnetization shown in Fig. \ref{fig6}(a) remains constant over most of the pressure range except for an abrupt drop from its saturated value to zero, which emerges at a discontinuous pressure-induced phase transition at $p \approx 0.028$. Increasing temperature reduces the magnetization jump and shifts it toward lower pressures as demonstrated by the isothermal dependence at $k_{\rm B}T/J_0 = 0.03$. At the critical temperature $k_{\rm B}T_c/J_0 \approx 0.048$, the discontinuity in the magnetization disappears and is replaced by an inflection point with a divergent slope, which corresponds to a critical point of the continuous phase transition shown by a star symbol positioned at $p=0$ in Fig. \ref{fig6}(a). For higher temperatures such as $k_{\rm B}T/J_0 = 0.06$, the magnetization varies smoothly with pressure across the entire pressure range. The magnetic susceptibility shown in Fig. \ref{fig6}(b) exhibits behavior fully consistent with this scenario. At low temperatures $k_{\rm B}T/J_0 \leq 0.048$, it displays a finite cusp associated with the discontinuous phase transition, which gradually evolves into a pronounced divergence as temperature approaches its critical value $k_{\rm B}T_c/J_0 \approx 0.048$. Above the critical temperature (e.g. $k_{\rm B}T/J_0 = 0.06$), the susceptibility displays a smooth continuous pressure dependence with a finite round maximum. The lattice response mirrors this magnetic behavior. Fig. \ref{fig6}(c) shows that increasing pressure induces at sufficiently low temperatures $k_{\rm B}T/J_0 \leq 0.048$ an abrupt compressible strain due to the discontinuous phase transition, which changes to a continuous compressible strain at and above the critical temperature $k_{\rm B}T/J_0 \gtrsim 0.048$. The pressure-induced changes of the inverse compressibility shown in Fig. \ref{fig6}(d) further corroborates the pronounced changes in the distortion parameter. The main panel and inset of Fig. \ref{fig6}(d) clearly demonstrate that the inverse compressibility exhibits small dip anomalies at the two lowest temperatures, while it completely diminishes at the critical point with the coordinates $p=0$ and $k_{\rm B}T_c/J_0 \approx 0.048$.

\begin{figure*}
\begin{center}
\includegraphics[width=0.5\textwidth]{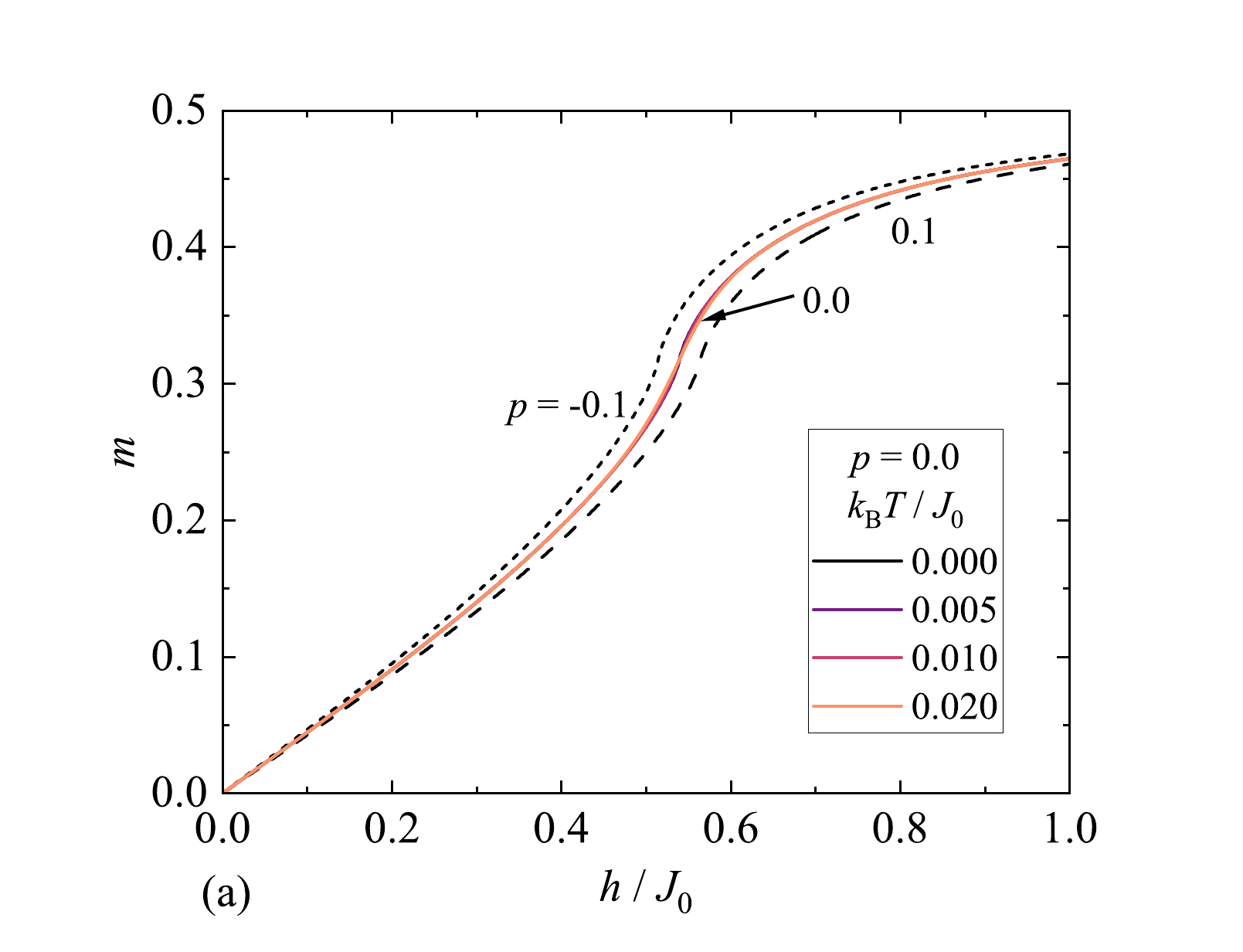}
\hspace*{-1cm}
\includegraphics[width=0.5\textwidth]{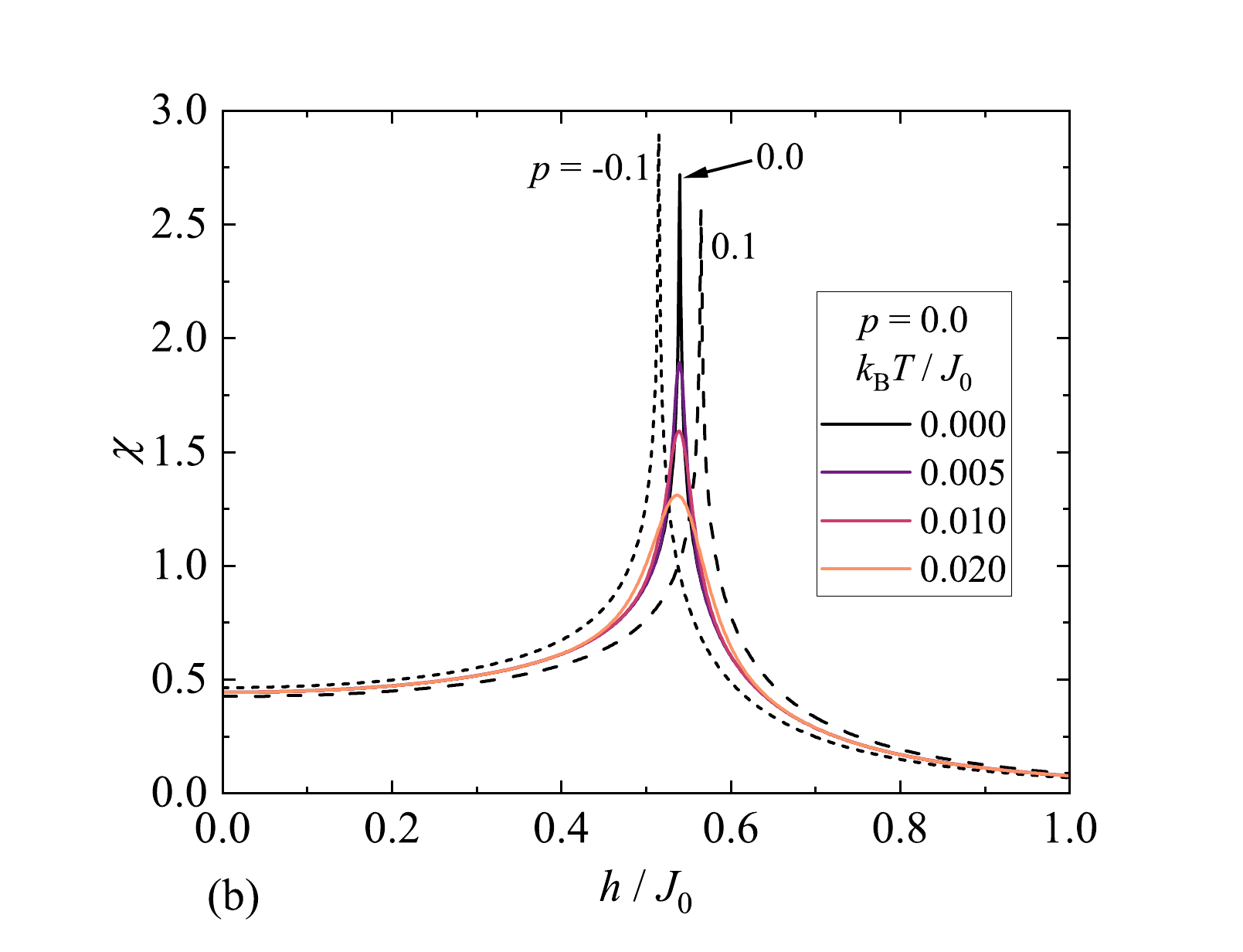}
\includegraphics[width=0.5\textwidth]{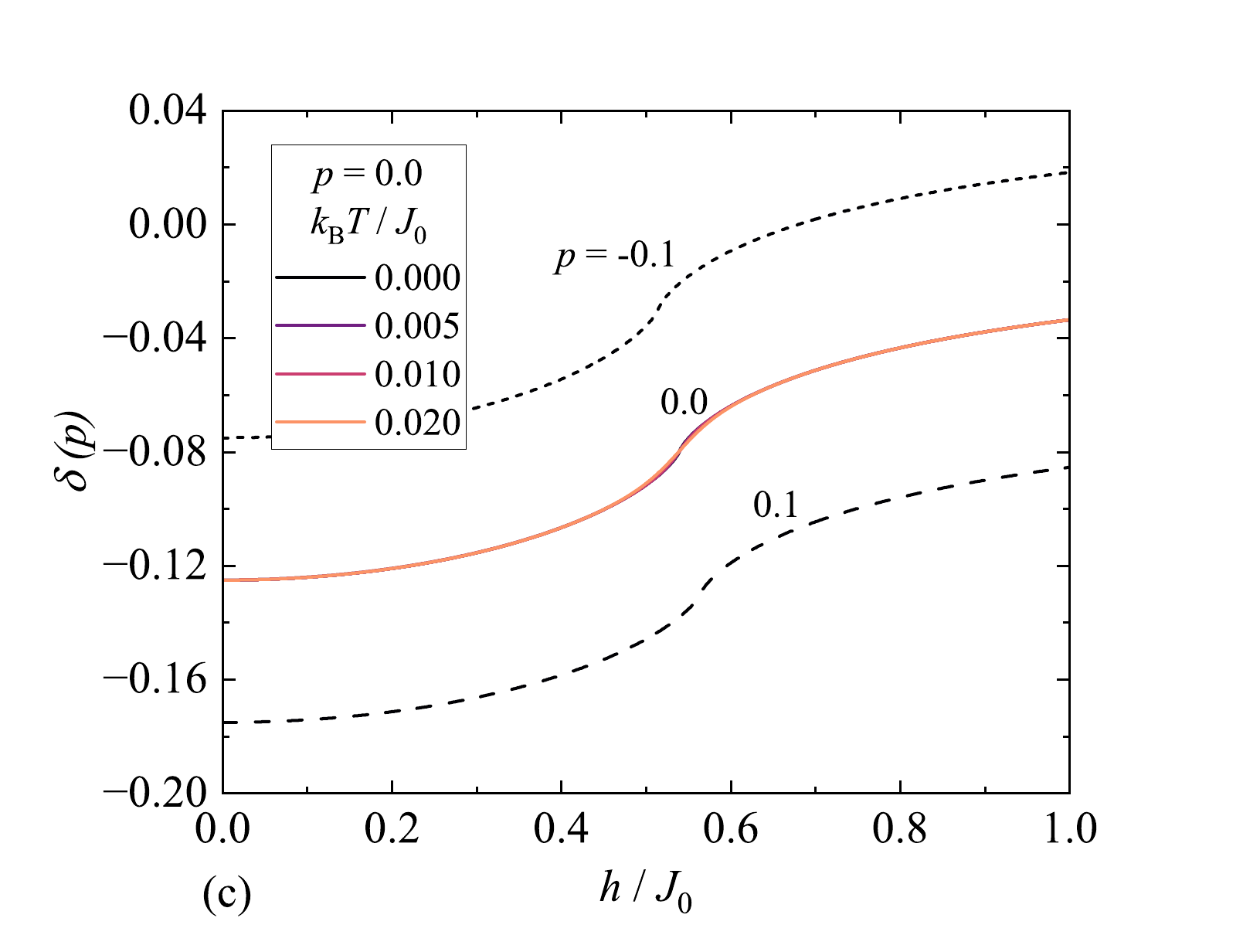}
\hspace*{-1cm}
\includegraphics[width=0.5\textwidth]{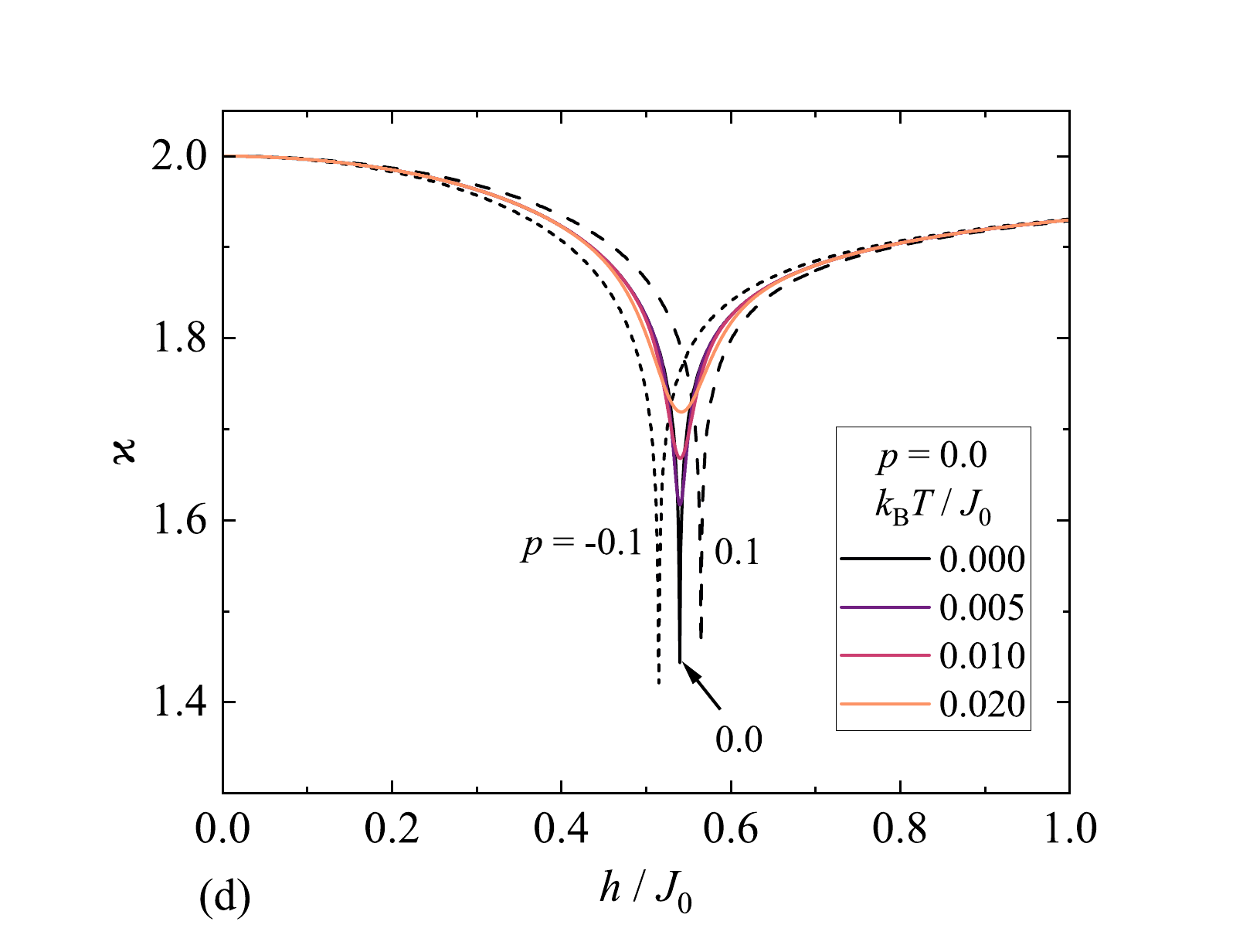}
\end{center}
\vspace{-0.9cm}
\caption{Magnetic-field dependence of (a) the magnetization, (b) the magnetic susceptibility, (c) the distortion parameter, and (d) the inverse compressibility of the deformable spin-1/2 Ising chain in a transverse magnetic field at fixed pressure $p = 0$ and four distinct temperatures. Dashed and dotted black lines display additional results for zero temperature $k_{\rm B}T/J_0 = 0$ and two specific values of the nonzero pressure $p = \pm 0.1$.}
\label{fig7}       
\end{figure*}

\subsection*{B. Transverse field}
\label{Rsubsection_B}

Next, we examine the magnetic and elastic properties of the deformable spin-1/2 Ising chain in a transverse magnetic field. Figs. \ref{fig7}(a) and \ref{fig7}(b) show the field dependence of the magnetization and magnetic susceptibility at fixed pressure $p=0$ and four representative temperatures. The additional zero-temperature data for two specific cases with nonzero pressure $p = \pm 0.1$ are also included for comparison. For the zero-pressure case $p=0$, all low-temperature magnetization curves depicted in Fig. \ref{fig7}(a) are smooth, continuous, and nearly indistinguishable over the entire magnetic-field range. A closer inspection, however, reveals a nonanalytic singularity emerging in the zero-temperature magnetization curve at the critical field $h_c/J_0 \approx 0.54$, which provides a clear manifestation of the field-induced quantum phase transition. This quantum phase transition is more evident from Fig. \ref{fig7}(b), where the magnetic susceptibility exhibits at zero temperature a marked divergence near the critical field $h_c/J_0 \approx 0.54$. The magnetic-field variations of the magnetic susceptibility at any nonzero temperature is contrarily rounded off and replaced by a smooth round maximum, whose height is progressively suppressed and becomes broader with increasing temperature (see the curves for $k_{\rm B}T/J_0 = 0.005$, $0.01$, and $0.02$). This behavior reflects the thermal smearing of the quantum critical singularity. Evident signatures of field-driven quantum phase transitions can be also detected in the respective zero-temperature dependence of the magnetic susceptibility at nonzero pressure, which only shifts the critical field toward higher (lower) values for the compressible (tensile) strain $p>0$ ($p<0$) without qualitative altering the quantum critical behavior. 

The similar trends can be observed in the field dependence of the distortion parameter and inverse compressibility shown in Figs. \ref{fig7}(c) and \ref{fig7}(d) for the same set of parameters as used for the magnetic quantities. Fig. \ref{fig7}(c) demonstrates a continuous elongation of the chain with increasing transverse magnetic field at all considered temperatures, whereby all low-temperature curves appear nearly identical for a given pressure indicating that thermal effects remain weak in this regime. The field-driven phase transition occurs strictly at zero temperature as clearly manifested in the field variations of the inverse compressibility depicted in Fig. \ref{fig7}(d). A sharp dip in the inverse compressibility reflecting the singular elastic response associated with the quantum phase transition can be actually observed only at zero temperature. At finite temperatures, the dip-like anomaly is progressively rounded into a smooth minimum that becomes broader and shallower with increasing temperature (see the curves for $k_{\rm B}T/J_0 = 0.005$, $0.01$ and $0.02$). This thermal rounding mirrors the behavior observed in the magnetic susceptibility and confirms that the singular elastic anomaly is a genuine zero-temperature quantum critical effect.

It is worthwhile to remark that the anomalous behavior observed in the vicinity of the quantum phase transition originates from the singular behavior of the complete elliptic integrals of the first and second kind $E(r)$ and $K(r)$ (and their derivatives) when their modulus approaches the limiting value $r \approx 1$. According to Eq. (\ref{Zdeff}), the critical condition $J_0(1-\delta_c)-2h_c = 0$ can be expressed in terms of the critical value of the distortion parameter $\delta_c = -(p + J_0/2\pi)/\alpha$ determined by Eq. (\ref{TransZeroPaper1}). The zero-temperature field dependencies of all magnetic and elastic quantities accordingly exhibit singularities at the critical field $h_c/J_0 = 1/2 + (p_c + J_0/2\pi)/(2\alpha)$ in full agreement with the numerical results discussed above. This analytical condition therefore provides a consistent explanation of the quantum critical singularities observed in the magnetization, susceptibility, distortion parameter, and inverse compressibility.

\begin{figure*}
\begin{center}
\includegraphics[width=0.5\textwidth]{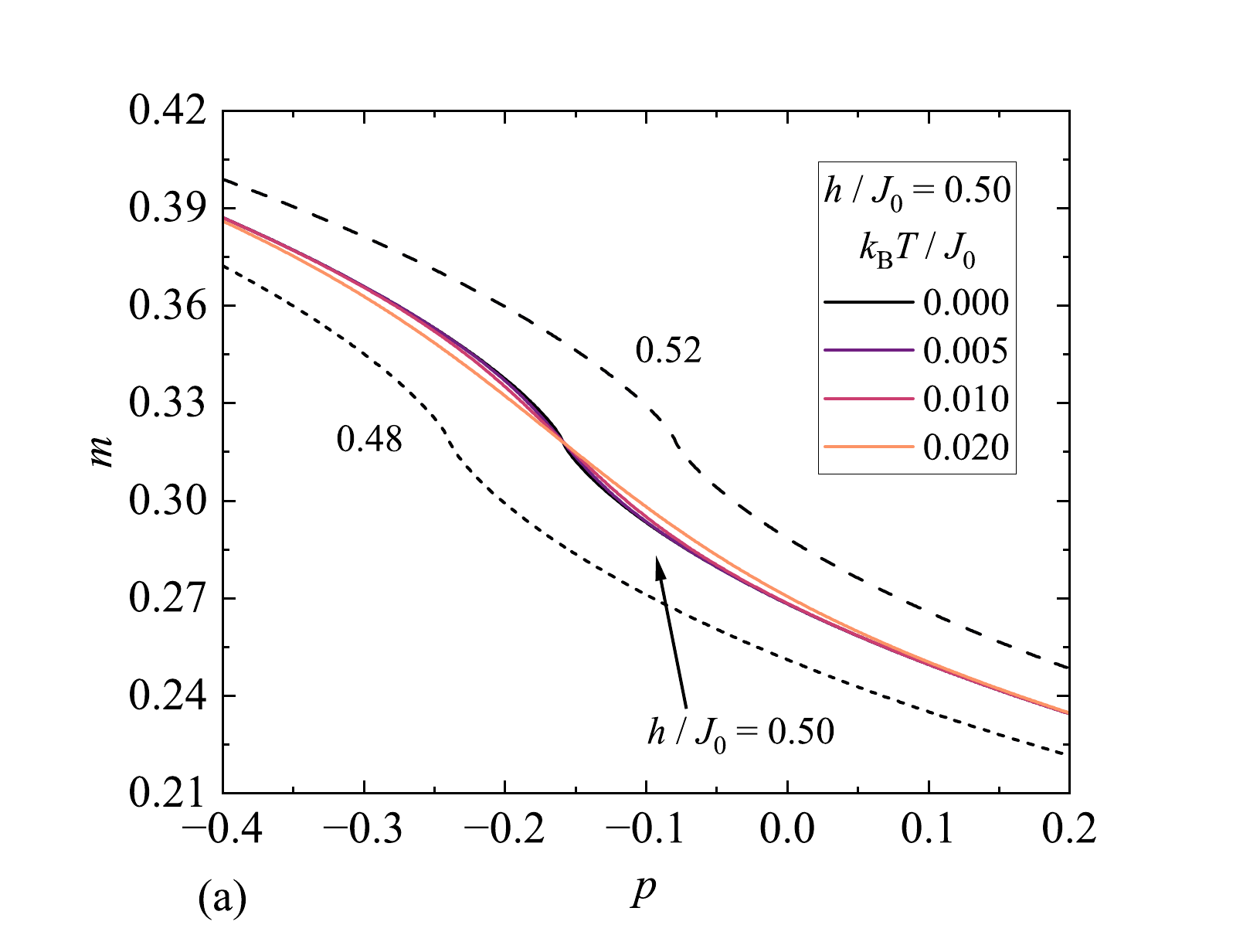}
\hspace*{-1cm}
\includegraphics[width=0.5\textwidth]{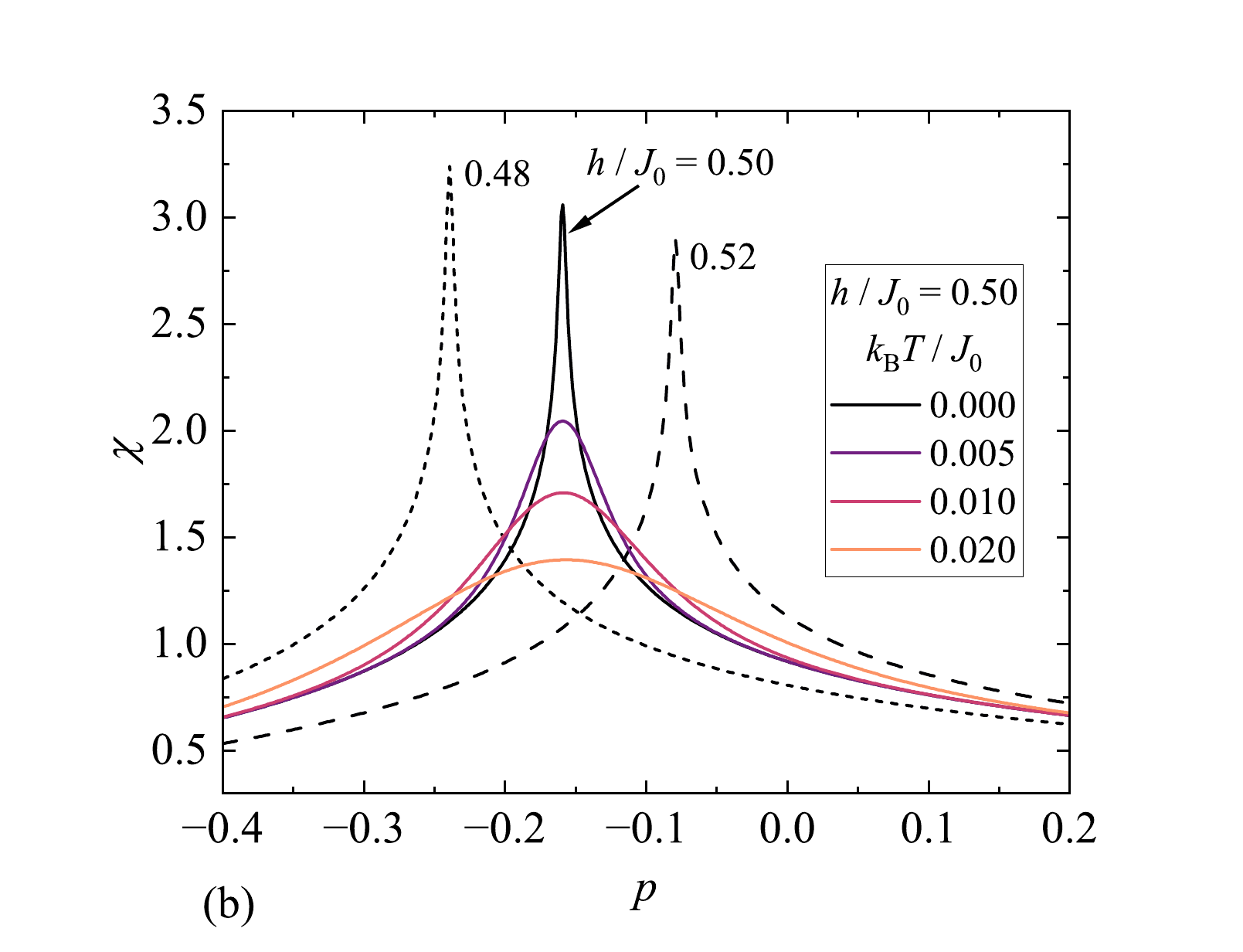}
\includegraphics[width=0.5\textwidth]{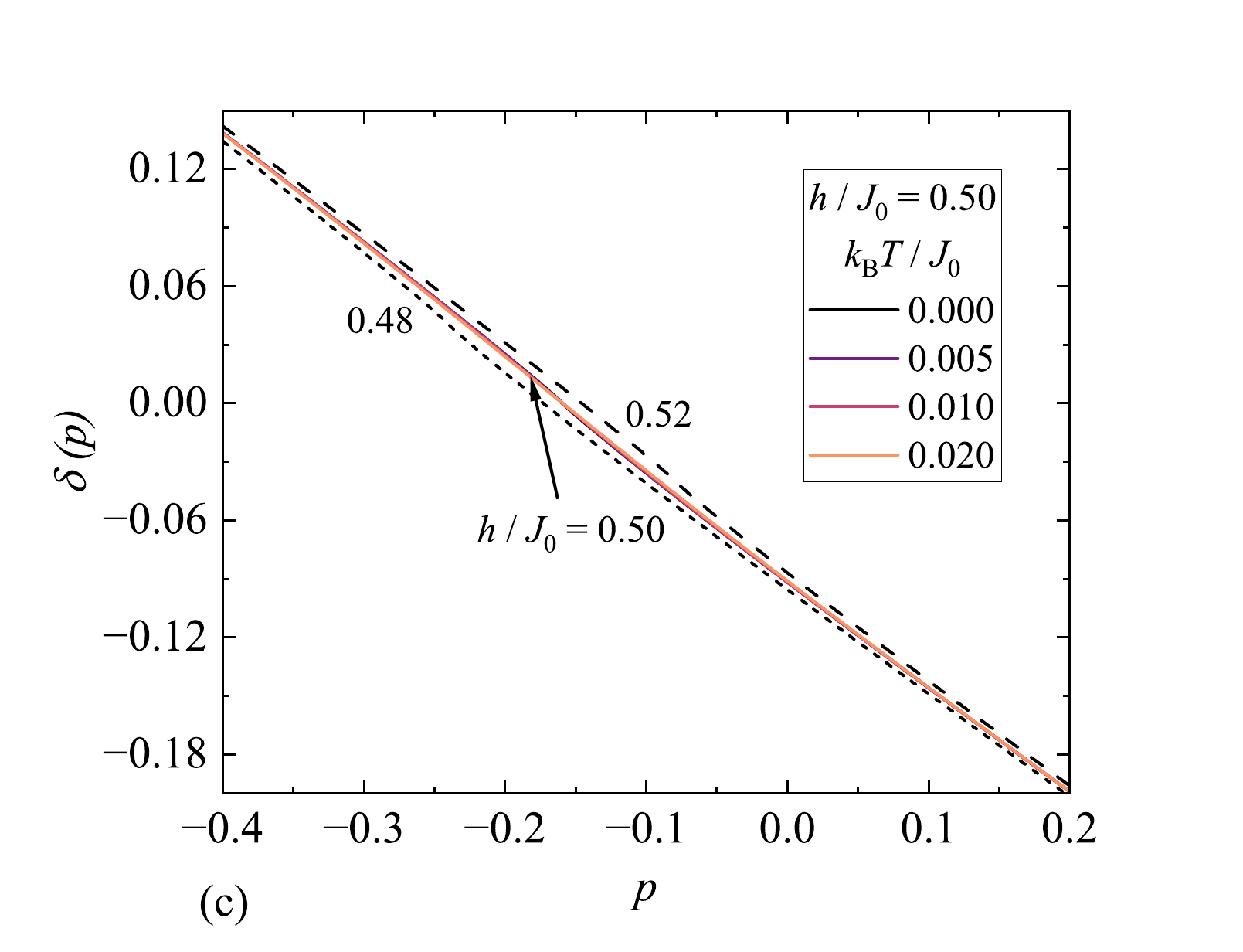}
\hspace*{-1cm}
\includegraphics[width=0.5\textwidth]{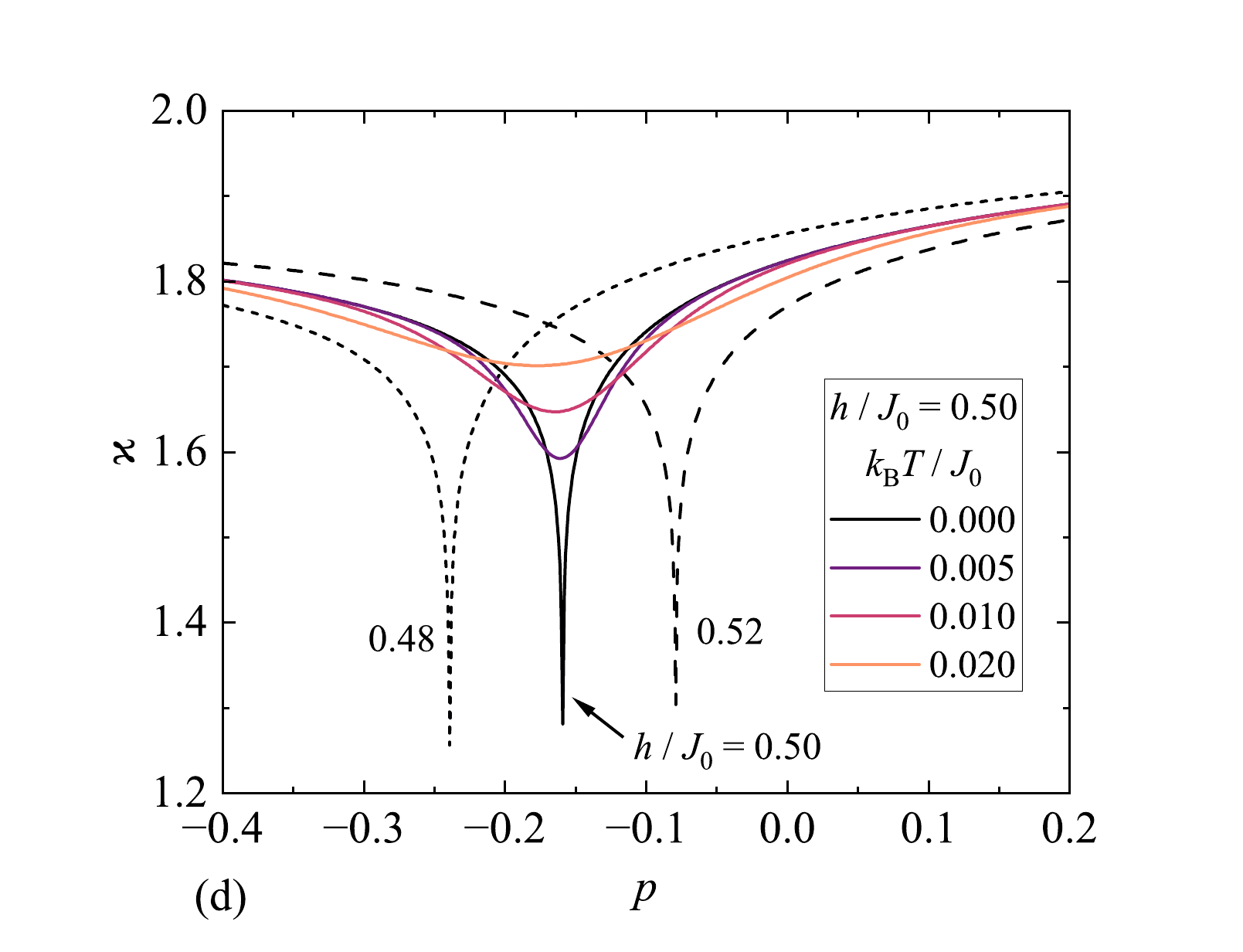}
\end{center}
\vspace{-0.9cm}
\caption{Pressure dependence of (a) the magnetization, (b) the magnetic susceptibility, (c) the distortion parameter, and (d) the inverse compressibility of the deformable spin-1/2 Ising chain in a transverse magnetic field $h/J_0 = 0.5$ and four distinct temperatures. Dashed and dotted black lines display additional zero-temperature results ($k_{\rm B}T/J_0 = 0$) for two other magnetic fields $h/J_0 = 0.48$ and $0.52$.}
\label{fig8}       
\end{figure*}

The deformable spin-1/2 Ising chain under a fixed transverse magnetic field can also undergo a quantum phase transition driven by pressure tuning. The corresponding critical pressure $p_c = 2\alpha(h/J_0 - 1/2) - J_0/2\pi$, at which this quantum phase transition occurs, is obtained analogously to the field-induced quantum critical condition discussed above. Figs. \ref{fig8}(a) and \ref{fig8}(b) display the pressure dependence of the magnetization and magnetic susceptibility at the fixed magnetic field $h/J_0 = 0.5$ and four representative temperatures. Additional zero-temperature results for other two field strengths $h/J_0 = 0.48$ and $0.52$ are also included to illustrate how generic the observed pressure-induced quantum phase transition is. Fig. \ref{fig8}(a) demonstrates that the magnetization decreases continuously with increasing pressure, whereas all low-temperature curves at a given pressure nearly coincide with only a minor quantitative differences. The zero-temperature singularity of the magnetization can be located at the critical pressure $p_c \approx -0.16$, which corresponds to the crossing point for the pressure variations of the magnetization at low temperatures. A sharp cusp in the magnetic susceptibility presented in Fig. \ref{fig8}(b) provides a more pronounced manifestation of the pressure-induced quantum phase transition appearing at zero temperature and the critical pressure. At finite temperatures, the singular cusp of the susceptibility is rounded into a smooth maximum, whose height is gradually suppressed and width increases with increasing temperature reflecting thermal smearing of the quantum critical singularity. The pressure dependence of the distortion parameter plotted in Fig. \ref{fig8}(c) implies that the lattice shrinks quasi-linearly with increasing pressure, whereby the most pronounced deviation from the quasi-linear dependence appears near the quantum critical point. A sharp dip singularity in the pressure dependence of the inverse compressibility shown in Fig. \ref{fig8}(d) bears further evidence of the pressure-induced quantum phase transition emerging at the critical pressure. At finite temperatures, the sharp dip evolves into a round minimum, which becomes broader and flatter as temperature increases. This observation again confirms that the pressure-induced quantum phase transition is confined to absolute zero temperature.

\begin{figure*}
\begin{center}
\includegraphics[width=0.5\textwidth]{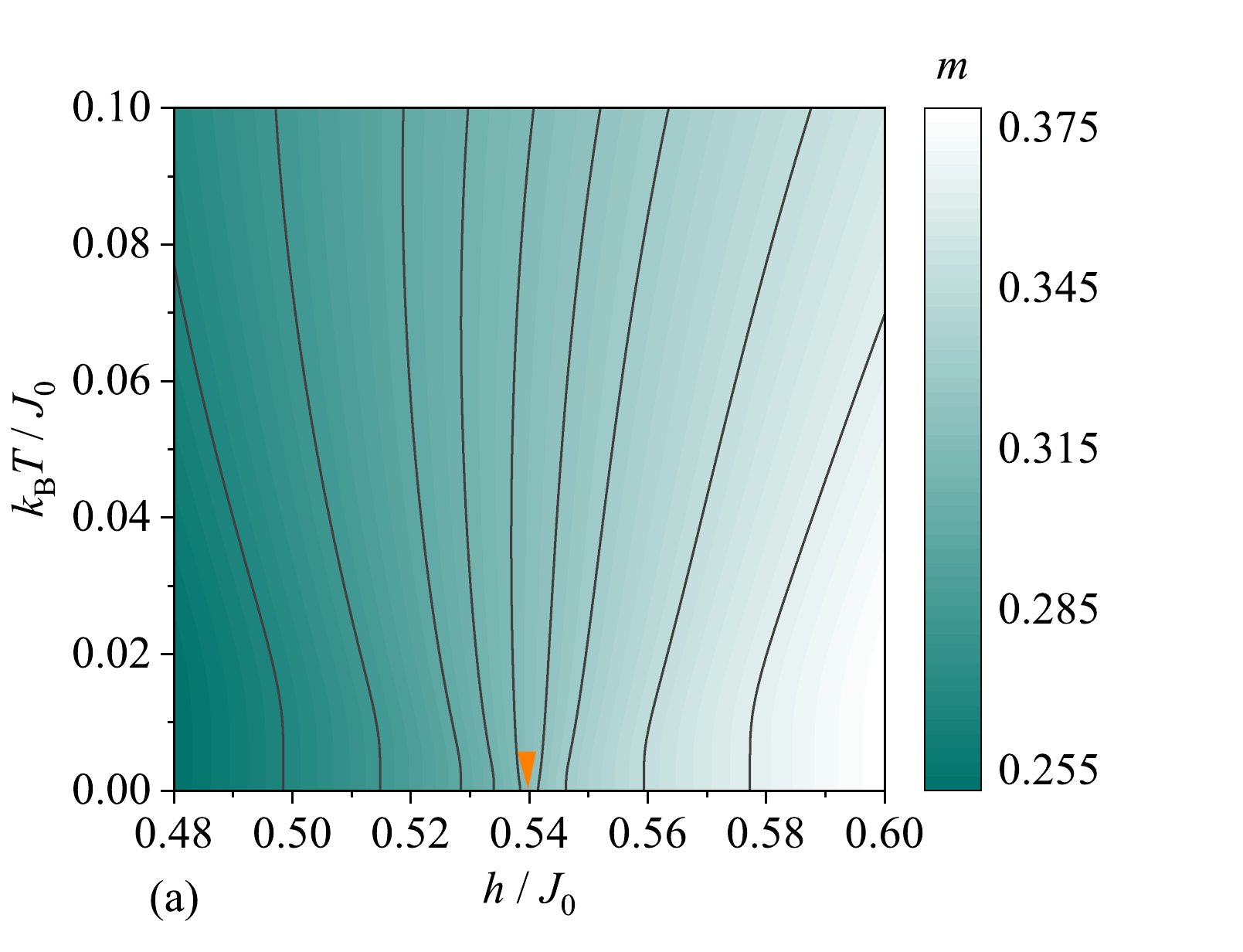}
\hspace*{-1cm}
\includegraphics[width=0.5\textwidth]{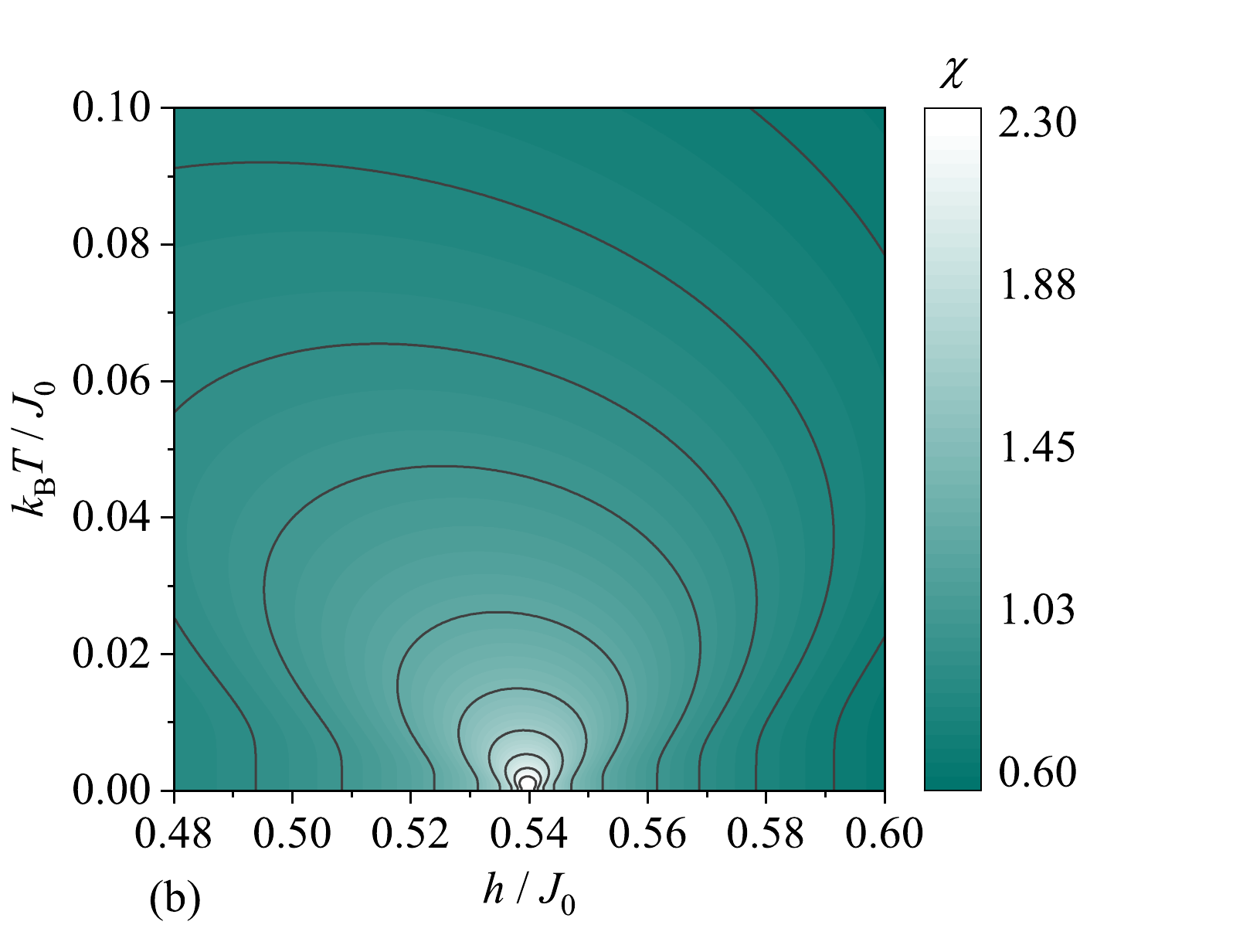}
\includegraphics[width=0.5\textwidth]{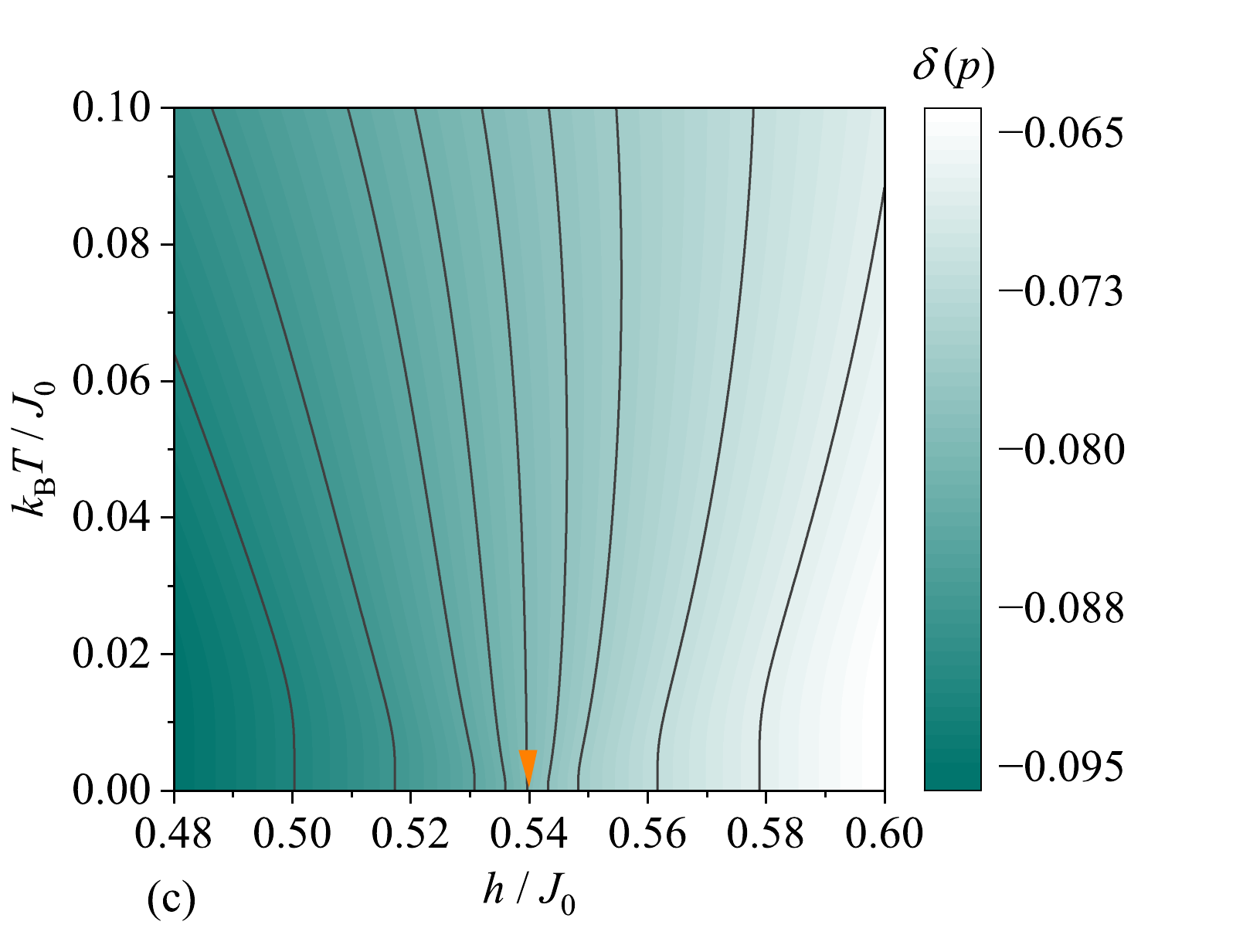}
\hspace*{-1cm}
\includegraphics[width=0.5\textwidth]{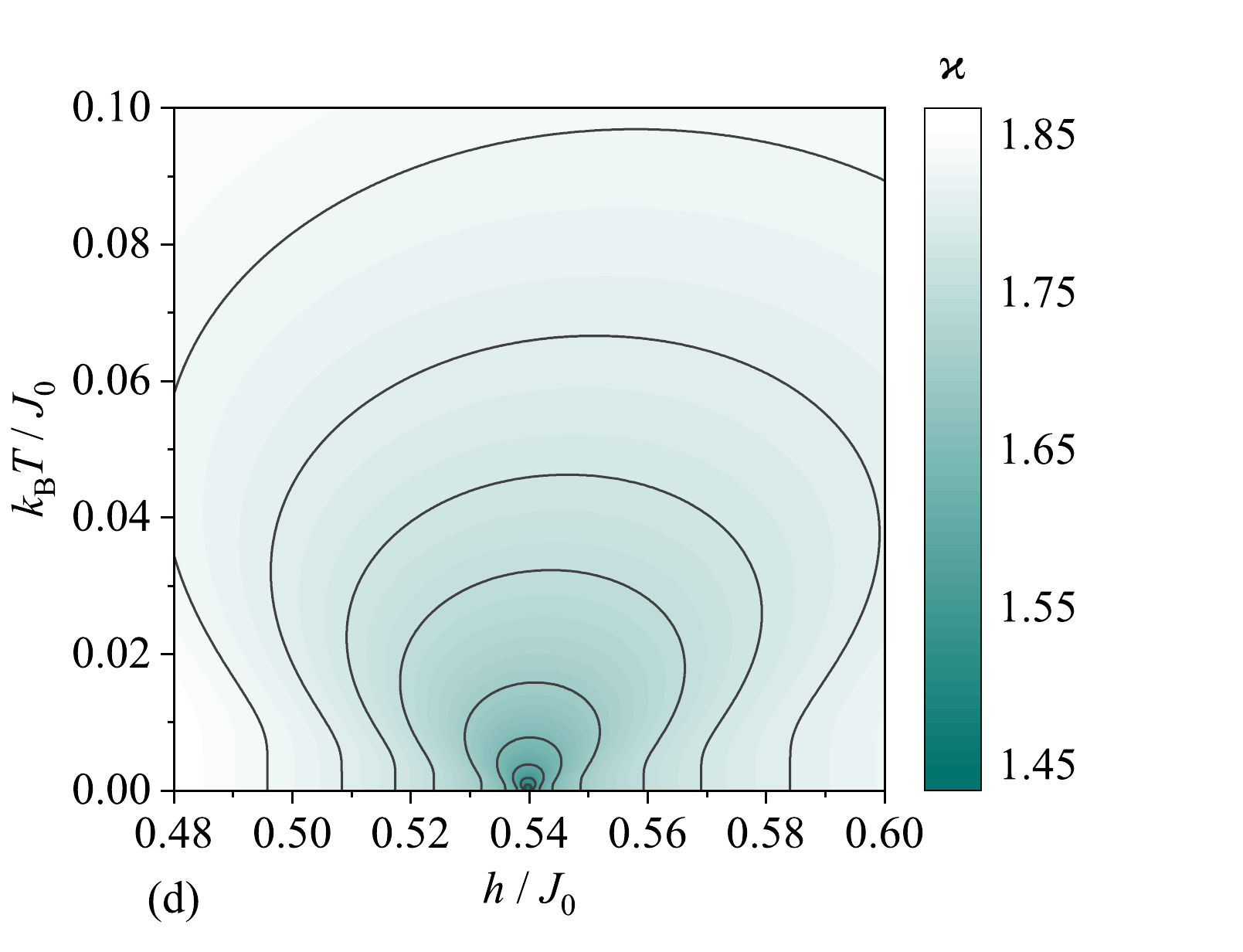}
\end{center}
\vspace{-0.9cm}
\caption{Density plots of (a) the magnetization, (b) the magnetic susceptibility, (c) the distortion parameter, and (d) the inverse compressibility of the deformable spin-1/2 Ising chain in a transverse magnetic field shown in the field-temperature plane at zero pressure $p=0$. Orange arrow indicates the quantum critical point.}
\label{fig9}       
\end{figure*}

To complete the overall picture, Figs. \ref{fig9}(a) and \ref{fig9}(b) present density plots of the magnetization and magnetic susceptibility of the deformable spin-1/2 Ising chain in a transverse magnetic field in the field-temperature plane at zero pressure $p=0$. The most prominent feature of the magnetization density plot in Fig. \ref{fig9}(a) is increasing density of contour lines in the vicinity of the critical field $h_c/J_0 \approx 0.54$ reflecting the enhanced field sensitivity near the quantum critical point. A similar accumulation of contour lines is also observed in the density plot of the distortion parameter illustrated in Fig. \ref{fig9}(c) indicating that the elastic response also becomes strongly field dependent in the same critical region. On the other hand, the density plots of the magnetic susceptibility and inverse compressibility in Figs.~\ref{fig9}(b) and \ref{fig9}(d) exhibit nested arc-shaped contour lines centered around the critical field $h_c/J_0 \approx 0.54$. 
These contour lines thus highlight the pronounced extremal behavior of the corresponding response functions near the quantum critical point. While the arc-shaped contour lines surround a global maximum in the magnetic susceptibility, they enclose a global minimum in the inverse compressibility reflecting the critical softening of the lattice accompanying the quantum phase transition.

\begin{figure*}
\begin{center}
\includegraphics[width=0.5\textwidth]{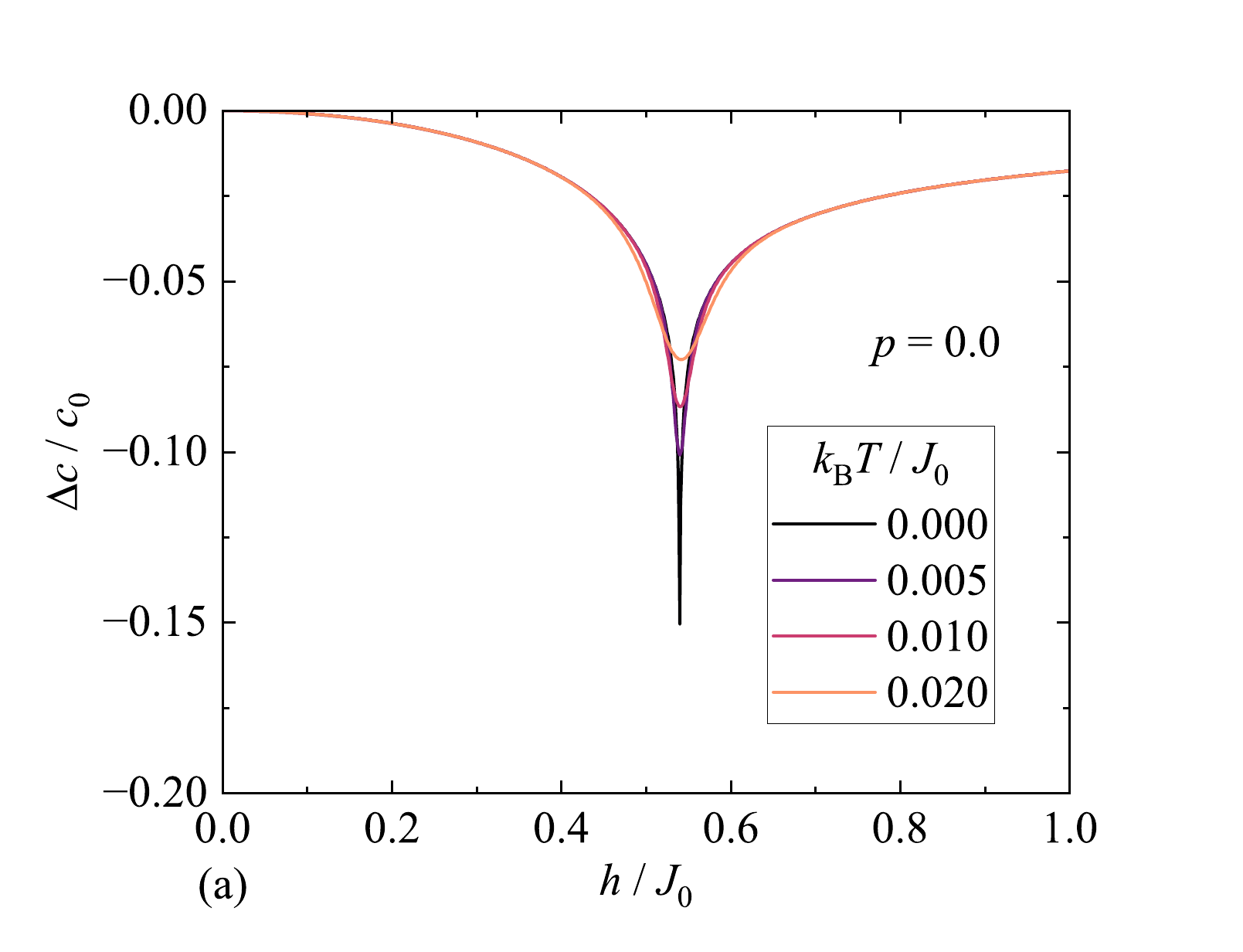}
\hspace*{-1cm}
\includegraphics[width=0.5\textwidth]{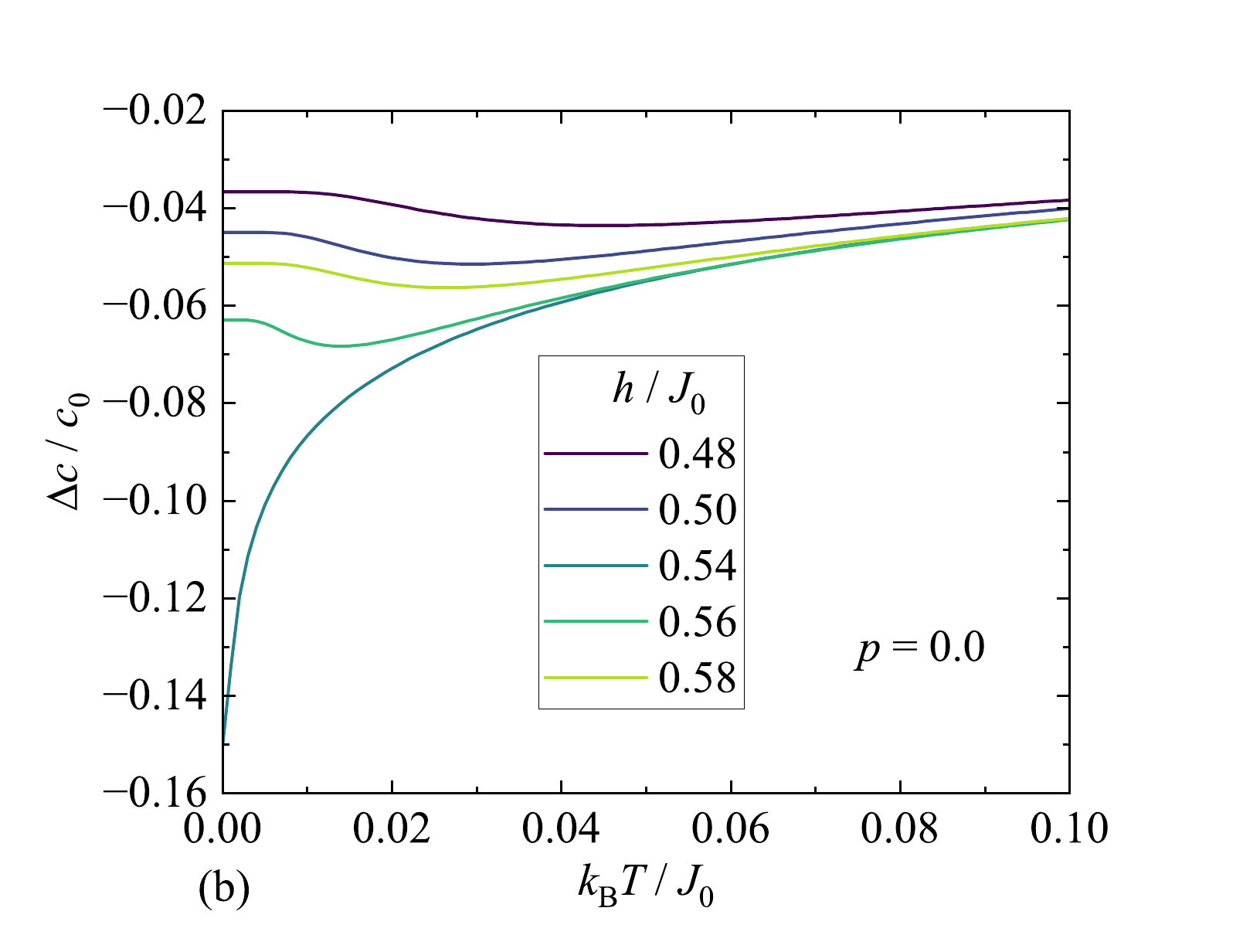}
\end{center}
\vspace{-0.9cm}
\caption{Magnetic-field (a) and temperature (b) dependence of the relative change in the sound velocity of the deformable spin-1/2 Ising chain in a transverse magnetic field at fixed zero pressure $p = 0$.}
\label{fig10}       
\end{figure*}

A change in the lattice spacing in the deformable spin-1/2 Ising chain subjected to a transverse magnetic field directly influences the sound velocity via the magnetoelastic coupling. Fig. \ref{fig10}(a) displays the field dependence of the relative change in the sound velocity, which qualitatively reflects the behavior of the inverse compressibility discussed above. In particular, the relative change of the sound velocity diminishes at zero temperature as the critical field $h_c/J_0 \approx 0.5$  is approached. At finite temperatures, the sharp dip-like singularity in the relative change of the sound velocity is rouned into a smooth finite minimum whose depth gradually reduces with increasing temperature. The sound attenuation due to the magnetoelastic coupling is also reflected in a marked temperature dependence demonstrated in Fig. \ref{fig10}(b) for a few selected values of the external magnetic field. A relatively broad and shallow minimum of the relative change of the sound velocity appears for a subcritical field $h/J_0 = 0.48$ near $k_{\rm B}T/J_0 \approx 0.045$, while this minimum becomes more pronounced and progressively shifts toward lower temperatures upon strengthening of the transverse magnetic field (e.g., $h/J_0 = 0.5$). When the magnetic field is fixed at its critical value $h_c/J_0 \approx 0.54$ for zero pressure $p = 0$, the relative change in the sound velocity decreases monotonically upon lowering the temperature and approaches its global minimum when reaching the quantum critical point at zero temperature. This behavior confirms that complete lattice softening occurs strictly at the quantum critical point.

\section{Concluding remarks}
\label{conclusion}

In the present article, we have systematically investigated the magnetization, magnetic susceptibility, distortion parameter, inverse compressibility, and the relative change in the sound velocity of the deformable spin-1/2 Ising chain subjected to either a longitudinal or a transverse magnetic field. Within the static approximation of the adiabatic regime, the nearest-neighbor exchange interaction was assumed to depend linearly on a uniform lattice distortion parameter $J = J_0(1 - |\kappa|\delta)$ thereby incorporating magnetoelastic coupling at the microscopic level. For both field orientations, exact expressions for the variational Gibbs free energy were employed and subsequently minimized with respect to the distortion parameter in a fully self-consistent manner. 

It has been demonstrated that the deformable spin-1/2 Ising chain in the longitudinal magnetic field undergoes a discontinuous thermal phase transition at sufficiently low temperatures, which can be driven by tuning the magnetic field, pressure, or temperature. The resulting line of discontinuous thermal phase transitions terminates at a critical point associated with the continuous thermal phase transition. The discontinuous transitions are accompanied by metastable states originating from multiple local minima of the variational Gibbs free energy. As a consequence, a magnetic hysteresis emerges in the low-temperature magnetization process under quasi-static field sweeps. The discontinuous thermal phase transitions are most clearly manifested in the cusp- and dip-like singularities of the magnetic susceptibility and inverse compressibility. In contrast, the continuous thermal phase transition can be most recognized through a diverging magnetic susceptibility and a vanishing inverse compressibility signaling critical magnetic behavior and elastic softening, respectively. 

Compared to this, we have furnished a rigorous evidence that the deformable spin-1/2 Ising chain in a transverse magnetic field undergoes exclusively a continuous phase transition at absolute zero temperature, which is driven either upon varying of the external magnetic field or pressure. Consequently, the deformable spin-1/2 Ising chain in the transverse magnetic field cannot exhibit a magnetic hysteresis due to a lack of metastable states that would be related to competing local minima of the variational Gibbs free energy. Although the quantum phase transition cannot be experimentally reached because the fundamental laws of thermodynamics forbid reaching of absolute zero temperature, the pronounced low-temperature features of the magnetic susceptibility and inverse compressibility still provide clear indirect signatures of the underlying quantum critical point. Moreover, the vigorous variation of the relative change in the sound velocity provides an alternative experimentally accessible probe of both quantum and thermal phase transitions induced by the magnetoelastic coupling. 

Last but not least, the theoretical framework developed in this work can be straightforwardly adapted to other exactly solvable one-dimensional quantum spin systems such as the spin-1/2 XX chain in a transverse magnetic field, various Ising-Heisenberg spin chains, or even more general one-dimensional quantum spin models where exact solutions remain elusive. 

\begin{acknowledgments}
The authors acknowledge funding by the grant of Slovak Research and Development Agency under the contract No. APVV-24-0091 and by the grant of The Ministry of Education, Research, Development and Youth of the Slovak Republic under the contract No. VEGA 1/0298/25.
\end{acknowledgments}

\end{document}